\documentclass[preprint,usenatbib,twocolumn]{aastex631}
\received{2024 July 17}
\revised{2024 October 8}
\accepted{2024 October 8}

\shortauthors{Arora et al.}

\newcommand{\vpec}{\ensuremath{\vec{v}_{\mathrm{pec}}}}

\newcommand{\mi}[1]{\textsf{m12i}}
\newcommand{\Msol}[1]{M\textsubscript{\(\odot\)}}

\newcommand{\upenn}{Department of Physics \& Astronomy, University of Pennsylvania, 209 S 33rd St, Philadelphia, PA 19104, USA}
\newcommand{\cca}{Center for Computational Astrophysics, Flatiron Institute, 162 5th Ave, New York, NY 10010, USA}

\shorttitle{Efficient halo orbit reconstructions}
\usepackage{courier}
\usepackage{multirow}
\usepackage{amsmath,amsfonts}
\usepackage[mathscr]{euscript}

\begin{document}

\title{Efficient and accurate force replay in cosmological-baryonic simulations}

\correspondingauthor{Arpit Arora}
\email{arora125@sas.upenn.edu}

\author[0000-0002-8354-7356]{Arpit Arora}
\affiliation{\upenn}
\affiliation{Department of Astronomy, University of Washington, Seattle, WA 98195, USA }

\author[0000-0003-3939-3297]{Robyn Sanderson}
\affiliation{\upenn}

\author{Christopher Regan}
\affiliation{\upenn}

\author[0000-0001-7107-1744]{Nicol\'as Garavito-Camargo}
\affiliation{\cca}

\author[0000-0003-3792-8665]{Emily Bregou}
\affiliation{\upenn}
\affiliation{Institut de Ciencies del Cosmos (ICCUB), Universitat de Barcelona (IEEC-UB), Mart\'i i Franques 1, E-08028 Barcelona, Spain}

\author[0000-0001-5214-8822]{Nondh Panithanpaisal}
\affiliation{\upenn}
\affiliation{Carnegie Observatories, 813 Santa Barbara St, Pasadena, CA 91101, USA}
\affiliation{TAPIR, California Institute of Technology, Pasadena, CA 91125, USA}

\author[0000-0003-0603-8942]{Andrew Wetzel}
\affiliation{Department of Physics \& Astronomy, University of California, Davis, CA 95616, USA}

\author[0000-0002-6993-0826]{Emily C. Cunningham}
\altaffiliation{NASA Hubble Fellow}
\affiliation{Department of Astronomy, Columbia University, 550 West 120th Street, New York, NY 10027, USA}

\author[0000-0003-3217-5967]{Sarah R. Loebman}
\affiliation{Department of Physics, University of California, Merced, 5200 Lake Road, Merced, CA 95343, USA}

\author[0000-0002-7352-6252]{Adriana Dropulic}
\affiliation{Department of Physics, Princeton University, Princeton, NJ 08544, USA }

\author[0000-0003-2497-091X]{Nora Shipp}
\affiliation{Department of Astronomy, University of Washington, Seattle, WA 98195, USA }
\affiliation{McWilliams Center for Cosmology, Department of Physics, Carnegie Mellon University, Pittsburgh, PA 15213, USA }

\begin{abstract}
We construct time-evolving gravitational potential models for a Milky Way-mass galaxy from the FIRE-2 suite of cosmological-baryonic simulations using basis function expansions. These models capture the angular variation with spherical harmonics for the halo and azimuthal harmonics for the disk, and the radial or meridional plane variation with splines. We fit low-order expansions (4 angular/harmonic terms) to the galaxy's potential for each snapshot, spaced roughly 25 Myr apart, over the last 4 Gyr of its evolution, then extract the forces at discrete times and interpolate them between adjacent snapshots for forward orbit integration. Our method reconstructs the forces felt by simulation particles with high fidelity, with 95\% of both stars and dark matter, outside of self-gravitating subhalos, exhibiting errors $\leq4\%$ in both the disk and the halo. Imposing symmetry on the model systematically increases these errors, particularly for disk particles, which show greater sensitivity to imposed symmetries. The majority of orbits recovered using the models exhibit positional errors $\leq10\%$ for 2-3 orbital periods, with higher errors for orbits that spend more time near the galactic center. Approximate integrals of motion are retrieved with high accuracy even with a larger potential sampling interval of 200 Myr. After 4 Gyr of integration, 43\% and 70\% of orbits have total energy and angular momentum errors within 10\%, respectively. Consequently, there is higher reliability in orbital shape parameters such as pericenters and apocenters, with errors $\sim10\%$ even after multiple orbital periods. These techniques have diverse applications, including studying satellite disruption in cosmological contexts. 
\end{abstract}

\section{Introduction} \label{sec:intro}

In the Cold Dark Matter (CDM) paradigm, dark matter (DM) halos grow hierarchically by accreting mass inside the cosmic web. The structure of these DM halos in CDM has been found to follow `universal' power law density and potential profiles in DM-only simulations \citep{einasto1965construction, Navarro_1997}, with some variations in baryonic simulations \citep[e.g.][]{lazar2020dark}. These halo profiles have enabled the study of the properties of many physical phenomena in the Universe, such as gravitational lenses and the internal dynamics of stars within a halo.

As the halos assemble over time, their internal structure evolves, such as their concentration \citep{jing1998confronting, wechsler2002concentrations, diemand2007evolution} and shape \citep{vera2013constraints, prada2019dark, emami2021morphological}. Furthermore, the evolution of the density profile is even more complex in simulations that include baryonic matter, as baryonic processes such as gas cooling, star formation, and feedback mechanisms can effectively add and redistribute the mass in the system \citep{wetzel2015physical, lazar2020dark, santistevan2024modelling}. Additionally, halos can undergo significant satellite mergers, such as those involving the LMC \citep{besla2007magellanic} and progenitor of the Sagittarius stream \citep{johnston1995disruption} in the Milky Way (MW), which break symmetry and induce disequilibrium in both the halo \citep{petersen2020reflex, cunningham2020quantifying, vasiliev2021tango, garavito2021quantifying,  vasiliev2023dear} and the disk \citep{laporte2018influence, hunt2019signatures,  vasiliev2021tango}. 

Traditional orbit modeling techniques utilize static, symmetric halo models, which provide a simplified and computationally efficient means of analyzing orbits. Generally, they decompose the MW into three or more components -- the bulge, the disk, and the halo -- and model the contribution of each component separately. The bulge and the halo are generally represented by the Navarro–Frenk–White potential model (spherical or flattened) \citep{Navarro_1997}, whereas the Miyamoto-Nagai potential model is commonly assumed for the disk \citep{MN1975}. The parameters of these potential models are then fit to the available dynamical data of the MW \citep[e.g.,][]{bovy2013, mcmillan2017} and used to backward integrate the orbits of dwarf galaxies \citep[e.g.,][]{patel2016orbits, patel2018estimating, fritz2018gaia}, globular clusters \citep[e.g.,][]{pouliasis2017milky, massari2019origin, vasiliev2019proper}, or stellar streams \citep[e.g.,][]{bonaca2014milky, erkal2018modelling, malhan2019constraining, riley2020milky} using commonly available galactic dynamics tool such as \texttt{galpy} \citep{bovy2015galpy}, \texttt{gala} \citep{pw2017gala}, and \texttt{Agama} \citep{vasiliev2019agama}. Sometimes, these models include mass growth of the halo \citep[e.g.,][]{d2022uncertainties,ishchenko2023milky} and specific effects of perturbing bodies, such as the LMC \citep[e.g.,][]{lux2010determining, gomez2015and, erkal2019total, shipp2021measuring, pace2022proper, koposov2023s}. 

Recent studies have highlighted the limitations of static models in recovering orbital parameters within acceptable errors. By comparing integrated orbits with exact orbits in Milky Way-mass cosmological simulations \citep{d2022uncertainties, santistevan2024modelling}, and to time-evolving potential models \citep{sanders2020models, vasiliev2021tango}, researchers have observed significant discrepancies. \citet{santistevan2024modelling} noted an error of roughly 80\% for orbit shape parameters such as minimum pericenter and apocenter distances between recovered and true orbits of satellite dwarf galaxies integrated in a time-static model. Similarly, \citet{d2022uncertainties} found high errors in their integration scheme, which accounted for the mass growth of the potential but maintained a fixed shape. They quantified that the orbital parameter errors were comparable to those caused by a 30\% uncertainty in the host mass. Additionally, they observed that modeling a recent massive accretion event, such as the LMC, using a combination of two spherical parametric potentials, led to substantial errors in the recovered orbital parameters. These findings emphasize that capturing the time-dependent evolution of halo and its response to massive mergers is particularly important when reconstructing the dynamics of the objects that reside in them, such as satellite galaxies, globular clusters, stellar streams, and stars \citep{wetzel2015satellite, d2021infall, d2022uncertainties, lilleengen2023effect, santistevan2023orbital, santistevan2024modelling}.

The evolution of the halo's structure can be captured by using Basis Function Expansions (BFE) models, which can accurately describe any arbitrary density field  \citep[e.g][]{weinberg1999adaptive, lowing2011halo, vasiliev2013new, sanders2020models, garavito2021quantifying, vasiliev2021tango, arora2022stability, petersen2022exp, lilleengen2023effect}. \citet{lowing2011halo} showed that by applying a Hernquist BFE (often referred to in the literature as a self-consistent field expansion) to every snapshot, spaced roughly 25 Myr, from a cosmological simulation of a DM halo and then interpolating the coefficients of the expansion in time allows substantial improvement in the orbital properties of a halo. Similarly, \citet{sanders2020models} showed high-fidelity orbit reconstruction in DM-only cosmological simulations of MW mass galaxies using compact BFE representation for the time-evolving halo potential, employing spherical harmonics for angular variation and bi-orthonormal basis functions or splines for radial variation. In this paper, we extend this methodology to model fully zoomed baryonic-cosmological simulations of MW mass galaxies from the Latte suite of the FIRE-2 project \citep{wetzel2016reconciling,wetzel2023public}. Unlike \citet{sanders2020models}, where only DM was considered, our simulations include stars and gas, necessitating adaptation of the BFE method with an azimuthal harmonic expansion to account for the baryonic component's flattened shape. Additionally, we utilize a fiducial temporal cadence with snapshots spaced at approximately 25 Myr intervals, and explore the effect of larger sampling intervals (up to 500 Myr) on the quality of recovered orbits. We demonstrate that sufficiently high fidelity halo orbit reconstruction can be achieved with our fiducial cadence, even though the potential in the inner regions can change much more rapidly. We assess the efficacy of this modified potential model in reconstructing orbits at a range of radii. We fit BFE to the potential at a series of discrete snapshots. Then, we approximate the time dependence by calculating a force/acceleration on a star at a given point in time as a linear interpolation of the forces in the two time-adjacent snapshots for orbit recovery. 

The paper is organized as follows:

In Sec.~\ref{sec:methods}, we provide a brief overview of the simulation (\S\ref{sec:sims}), potential modeling techniques (\S\ref{sec:pot_models}), and constraints on recovered particle forces in the potential model (\S\ref{sec:recon_forces_total}). In Sec.~\ref{sec:integ_sample}, we describe how we select orbits of \emph{halo} stars from the parent halo for reconstruction and integrate them in a cosmological setting. In Sec.~\ref{sec:recon_orbits}, we statistically quantify the quality of reconstructed orbits based on recovered 3D positions (\S\ref{sec:orbit_error_metric}), approximate integrals of motion, and orbital parameters such as pericenter and apocenter distances (\S\ref{sec:integ_motion}). We also evaluate the dependence of orbit quality on the sampling interval for potential models (\S\ref{sec:samp_interval}). In Sec.~\ref{sec:apps}, we demonstrate an application of the model by simulating the tidal disruption of a dwarf satellite. We discuss our findings and conclusions in Sec.~\ref{sec:disc_conc}.

\section{Methods} \label{sec:methods}

In this section, we detail our approach to reconstructing halo star orbits in a simulation of a MW-mass galaxy. We describe the galaxy simulation used (Sec.~\ref{sec:sims}), the BFE-based potential model fits to the simulation (Sec.~\ref{sec:pot_models}) and their force reconstruction for a sample of DM and star particles from the parent halo at present day (Sec.~\ref{sec:recon_forces_total}).   

\subsection{Simulations and coordinate system} \label{sec:sims}

We utilize a cosmological zoomed-in baryonic simulation of MW-mass galaxies from the \textit{Latte} suite \citep{wetzel2016reconciling} of the Feedback In Realistic Environments (FIRE) project, specifically \mi{}.\footnote{This simulation is publicly available \citep{wetzel2023public} at \url{http://flathub.flatironinstitute.org/fire}.} This simulations employ the FIRE-2 physics model \citep{Hopkins_2018} and is consistent with the $\Lambda \textrm{CDM}$ cosmology from Planck \citep{collaboration2015planck}: $\Omega_\Lambda=0.728$, $\Omega_\mathrm{matter}=0.272$, $\Omega_\mathrm{baryon}=0.0455$, $h=0.702$, $\sigma_8=0.807$, and $n_s=0.961$ \citep{wetzel2016reconciling}. \mi{} has a total mass of about $1.2 \times 10^{12}$ \Msol{} with a total stellar mass of $7 \times 10^{10}$ \Msol{} at present day, and initial star and gas particle masses of $m_\mathrm{b} = 7100$ \Msol{}, and DM particle mass $m_\mathrm{DM} = 35000$ \Msol{}. The high particle resolution enables the resolution of phase-space structures in the interstellar medium, facilitating the collapse of gas into well-resolved giant molecular clouds.

Snapshots are saved approximately every 25 Myr over the last 7 Gyr of the simulation; the most massive satellite merger in the last 6 Gyr has a total mass ratio of 1:45 relative to the MW \citep{arora2022stability, garavito2023co}. \citet{sanders2020models} demonstrated that reconstructed orbits in a DM-only simulation, with potential models sampled over intervals of 10 Myr and 40 Myr, yielded essentially identical results. Our sampling interval is sufficiently high to ensure high-fidelity orbit reconstructions for halo star orbits. We also show high fidelity can be achieved with a lower snapshot save rate of about 100 Myr. Additionally, the frequent snapshot intervals allow for the tracking of stellar orbits. 

These zoomed-in cosmological simulations are run in an arbitrary box frame on an expanding background with a non-zero total momentum. Initially, we recenter the simulation onto the host galaxy frame using the iterative shrinking spheres method \citep{power2003inner} to find a center-of-mass (COM) at each time step using star particles. This halo center in comoving coordinates is defined as $\vec{x}_\textrm{COM}(t) \equiv \vec{r}_\textrm{COM}(t)/a(t)$, where $a(t)$ is the cosmological scale factor, $\vec{r}_\textrm{COM}(t)$ is the physical COM position.  

Subsequently, we rotate all snapshots of the simulation to align the galactic disk with the XY plane at the present day--the principal axes. We define the host-centered rotation in the principal axes as the galactocentric frame. \citet{wang2020basis} and  \citet{arora2022stability} assessed the validity of this approach, considering the constancy of the disk's angular momentum over the past 7 Gyr of the simulations and the effectiveness of the potential modeling techniques in such systems. In \mi{}, the disk plane rotates by about 20 degrees over 7 Gyr up to the present day. This fixed rotation approximation also eliminates non-inertial forces associated with time-varying rotation (see Appendix.~\ref{app:NIF_exp_back}). While mergers with massive satellites such as the LMC \citep{besla2007magellanic} can affect the orientation of the disk \citep{baptista2023orientations}, our choice of \mi{}, {which features a quiescent merger history over the last 6 Gyr with the most massive merger being with a satellite of mass ratio 1:45}, ensures that the disk's orientation remains stable. {The models presented in this study are constructed to represent roughly the last 4 Gyr of the halo evolution, but they can be extended to any time frame as long as the disk remains oriented in the same direction}. However, careful consideration in modeling the disk is required if the orientation changes rapidly, as discussed in Appendix~\ref{app:NIF_exp_back}.

Force reconstruction for orbit integration must consider the acceleration of the comoving galactic center as a fictitious force within the galactocentric frame. Incorporating the non-inertial frame of the expanding cosmological background introduces another force.  The potential $\Phi(\vec{r},t)$ is modeled in the physical coordinates $\vec{r}$ of the non-inertial frame centered instantaneously on the galaxy. However, we define our equations of motion for the \emph{comoving} positions $\vec{x} \equiv \vec{r}/a(t)$ and the \emph{peculiar} velocities $\vpec \equiv a(t) \dot{\vec{x}}$. In these coordinates, the force acting on each particle for a fixed orientation of the disk is computed as (see Appendix \ref{app:NIF_exp_back}):
\begin{equation}
    \frac{d\vpec}{dt} = -\vec{\nabla}_{\vec{r}} \Phi(\vec{r},t)  - \frac{\dot{a} (t)}{a (t)} \vpec  - \frac{d \vec{u}_\textrm{COM}}{dt}
    \label{eq:force}
\end{equation}

It is useful to evaluate the relative contributions of the terms in Equation \ref{eq:force}. For a typical system, we have orbital velocities $\sim 100$s of km s$^{-1}$ and a Hubble parameter $\dot{a}/a \sim 1/14000$ {Myr}$^{-1}$, so the second term on the right is of order 0.01 km s$^{-1}$ Myr$^{-1}$. Meanwhile a typical value for $\vec{\nabla} \Phi$ at 30 kpc is roughly 1.5 km s$^{-1}$ Myr$^{-1}$, and $\dot{u}_\textrm{COM}$ is typically around 0.3-1 km s$^{-1}$ Myr$^{-1}$ \citep{arora2022stability}, so the contribution of the expanding background is relatively small for an isolated MW-mass halo, although it can be an order of magnitude larger for systems evolving in a Local Group environment, such as the MW-M31 system.

Following the approach of \citet{sanders2020models, arora2022stability, vasiliev2023dear}, we approximate $\dot{\vec{u}}_\textrm{COM}$ using second derivatives of smooth cubic spline fits to the halo's COM trajectory in comoving Cartesian coordinates.

\subsection{Potential Models} \label{sec:pot_models}

We employ the time-evolving low-order multipole potenital (TEMP) model introduced in \citet{arora2022stability}, fitted using a combination of BFE on the host density at each time step without assuming any symmetry conditions, constructed using \texttt{AGAMA} \citep{vasiliev2019agama}. {We fit these expansions to each snapshot in the final 4 Gyr of halo evolution to all the particles within 600 kpc of the galactic center}.These expansions are formulated as separable functions of 3D radial distance ($r$) or the cylindrical radius and height ($R$ and $Z$), and angular dependence, represented by orthogonal functions. 
To model the DM halo and hot gas ($T_{gas} \geq 10^{4.5}$ K), we use a spherical harmonic expansion in spherical coordinates to model the angular dependence ($\theta, \phi$), while the radial dependence ($r$) in the density is captured by evaluating the expansion coefficients in 25 logarithimically-spaced 1D radial grid nodes, interpolated using quintic splines \citep{vasiliev2013new}. The potential is written as   

\begin{equation}
    \Phi_\mathrm{halo}(r, \theta, \phi) = \sum_{\ell=0}^{
    \ell_\mathrm{max}
    } \sum_{m=-\ell}^{\ell} \Phi_{\ell m}(r) Y_{\ell}^m (\theta, \phi)
\end{equation}

We model the {stellar bulge} and flattened stellar and cold gas ($T_{gas} \leq 10^{4.5}$ K) component using a Fourier harmonic expansion in cylindrical coordinates ($R, \phi, Z$), where the expansion coefficients are computed on a 2D meridional plane ($R, Z$) with 25 and 40 grid nodes in $R$ and $Z$ respectively. The potential is written as           
\begin{equation}
    \Phi_\mathrm{disk} (R, \phi, Z) = \sum_{m=0}^{m_\mathrm{max}} \Phi_m (R, Z) e^{\iota m \phi}
\end{equation}

The methodology for fitting these models and computing the expansion coefficients ($\Phi_{\ell m}(r)$ and $\Phi_m (R, Z)$) is detailed in \citet{vasiliev2019agama}, and specifically applied to our simulations in \citet{arora2022stability}. The BFE adequately capture deformations in the disk \citep[e.g.][]{wang2020basis, petersen2022exp} and halo \citep[e.g.,][]{vasiliev2021tango, garavito2021quantifying} resulting from galactic evolution and satellite mergers. BFE has proven effective in reproducing orbits, even in the presence of massive satellites, in both idealized \citep{lilleengen2023effect, vasiliev2023dear} and cosmological simulations \citep{lowing2011halo, sanders2020models, arora2023lmc, donlon2024debris}. 

\subsection{Force reconstruction} \label{sec:recon_forces_total}
\begin{figure*}
\centering
\includegraphics[width=\textwidth,]{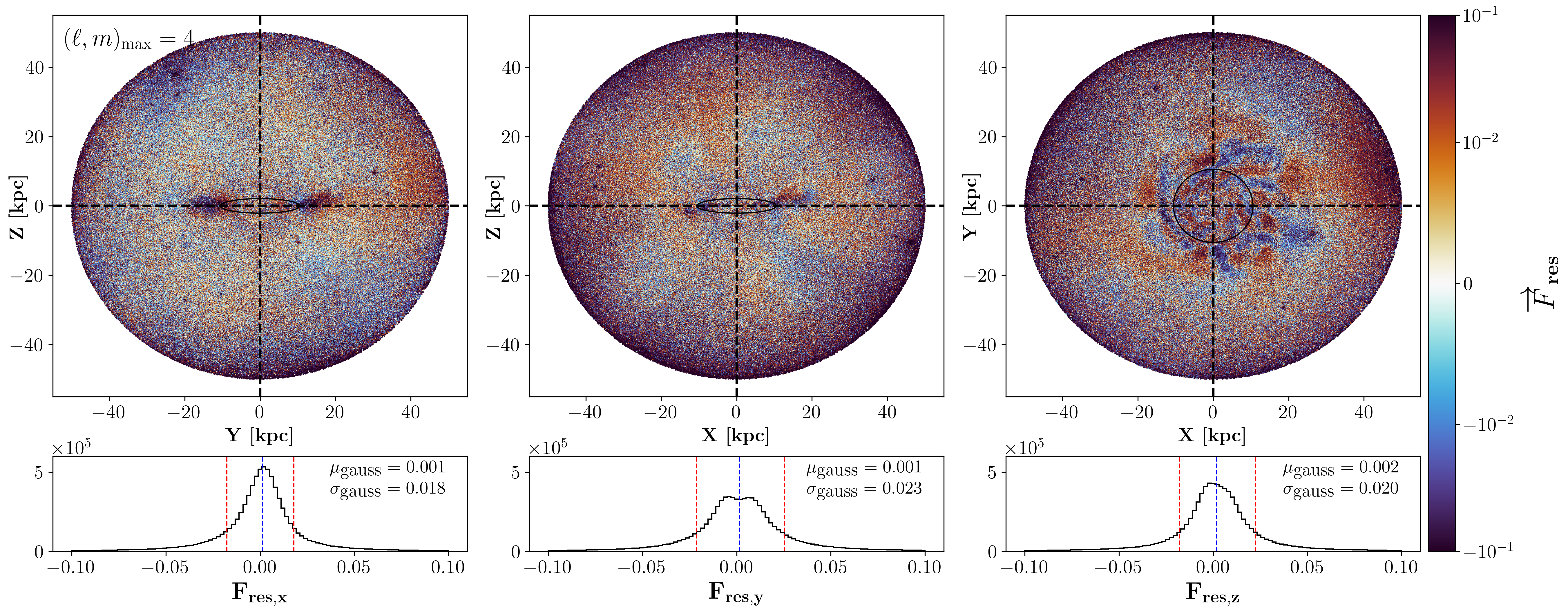}
\caption{{\bf Force reconstruction performance at present day.} {\it First row}: Relative force residual ($\vec{F}_\textrm{res} = {\vec{F}_\textrm{reconstructed}/\vec{F}_\textrm{true} -1}$) for DM particles within 50 kpc of the galactic center at the present day for the X direction in the YZ plane (left) the Y direction in the XZ plane (center), and the Z direction in the XY plane (right). The multipole model used to reconstruct the forces is truncated with maximum pole orders $\ell_\mathrm{max} = m_\mathrm{max} = 4$. The black ellipse encloses 90\% of the stellar mass in each panel. {\it Second row}: Histograms of the force residuals shown in each of the above panels. The 16th (red dashed line), 50th (blue dashed line), and 84th (red dashed line) quantiles, as well as the mean and standard deviation obtained from Gaussian fits, are marked for reference. Overall, the forces are reconstructed with under 10\% error, with over 95\% of the reconstructed forces having less than 4\% error. 
\label{fig:Fstd_n4n4}}
\end{figure*}

\setlength{\tabcolsep}{6pt}
\begin{table*}
\caption{{\bf Variation in performance with increasing pole order.} Force residual or the fractional variation from the true force distribution parameters for dark matter particles within 50 kpc from the galactic center at present day. Forces are reconstructed by assuming no symmetry in the model and varying the maximum pole order.}
\begin{center}
\begin{tabular}{|c|cc|cc|cc|cc|}
\hline
\multirow{2}{*}{$\mathbf{(\ell, m)_\textrm{max}}$} & \multicolumn{2}{c|}{$\mathbf{{F}_\textrm{res,X}}$}   & \multicolumn{2}{c|}{$\mathbf{{F}_\textrm{res,Y}}$}   & \multicolumn{2}{c|}{$\mathbf{{F}_\textrm{res,Z}}$}   & \multicolumn{2}{c|}{$\mathbf{{F}_\textrm{res, tot}}$}                                         \\ \cline{2-9} 
                                                   & $\mu \ [\cdot 10^{-2}]$ & $\sigma \ [\cdot 10^{-2}]$ & $\mu \ [\cdot 10^{-2}]$ & $\sigma \ [\cdot 10^{-2}]$ & $\mu \ [\cdot 10^{-2}]$ & $\sigma \ [\cdot 10^{-2}]$ & \multicolumn{1}{c}{$\mu \ [\cdot 10^{-2}]$} & \multicolumn{1}{c|}{$\sigma \ [\cdot 10^{-2}]$} \\ \hline
4                                                  & 0.13                    & 1.78                       & 0.14                    & 2.32                       & 0.16                    & 1.93                       & 0.17                                        & 0.84                                            \\
6                                                  & 0.13                    & 1.72                       & 0.13                    & 2.28                       & 0.17                    & 1.93                       & 0.17                                        & 0.83                                            \\
8                                                  & 0.14                    & 1.70                       & 0.13                    & 2.27                       & 0.17                    & 1.93                       & 0.17                                        & 0.83                                            \\
10                                                 & 0.13                    & 1.70                       & 0.13                    & 2.27                       & 0.16                    & 1.92                       & 0.17                                        & 0.83                                            \\ \hline
\end{tabular} \label{tab:force_res_order_l}
\end{center}
\raggedright
\textbf{Note.} $\mu$, $\sigma$ represent the mean and standard deviation, respectively, for each force residual or the fractional variation from the true force distribution in different Cartesian directions (as in bottom row of Figure \ref{fig:Fstd_n4n4}). $\mathbf{F}_\textrm{res, tot}$ denotes the residual on the total force magnitude. 
\end{table*}

Fig.~\ref{fig:Fstd_n4n4} shows the position-space distribution of the DM particles within 50 kpc of the galactic center at the present day where colors represent the force residual between reconstructed and true force, defined as $\vec{F}_\textrm{res} = {\vec{F}_\textrm{reconstructed}/\vec{F}_\textrm{true} -1}$ in each direction. In the first row, the left column illustrates the distribution in the YZ plane with the force residual in the X direction for each particle in the galactocentric frame and physical coordinates. The second and third columns depict force residuals in the Y and Z directions, respectively, in the XZ and XY planes. Our force reconstruction employs a truncated multipole model with maximum pole orders $\ell_\mathrm{max} = m_\mathrm{max} = 4$ in this case. The black ellipse encloses 90\% of the stellar mass in each panel. 

The second row shows the histograms of force residual distributions in each Cartesian coordinate, highlighting the 16th (red dashed line), 50th (blue dashed line), and 84th (red dashed line) quantiles, as well as the mean and standard deviation from Gaussian fits. The mean of all the distributions is close to zero, while the standard deviation represents a $\sim2\%$ error.   

Despite utilizing a low pole order of 4, the forces are reconstructed with under 10\% error for all the particles, with over 95\% of particles exhibiting less than 4\% error in reconstruction. Notably, discrepancies and higher errors ($\geq 5\%$) are predominantly observed in the baryonic disk, stemming from structures that are challenging to model with a low-order expansion, like spiral arms. With our focus on reproducing halo orbits in this work, we anticipate these errors in the inner regions of the galaxy to be negligible for most of our reconstructions. However, these errors can introduce biases in the reconstructed orbits, particularly for stars that spend a majority of their orbit in the inner regions (within 15 kpc of the galactic center). Moreover, the inner regions are susceptible to much more rapid changes in the potential compared to the halo, which our 25 Myr temporal cadence does not adequately capture. While limitations imposed by temporal cadence are noteworthy, we consider the bias in reconstructed forces to be a far more critical issue. Even higher azimuthal harmonic expansions, such as increasing to a pole order of 10, fail to address this primary concern. 

Table~\ref{tab:force_res_order_l} lists the mean ($\mu$) and standard deviation ($\sigma$) of force residuals for DM particles within 50 kpc from the galactic center across Cartesian directions ($\mathbf{{F}_\textrm{res, X}}, \mathbf{{F}_\textrm{res, Y}}$ and $\mathbf{{F}_\textrm{res, Z}}$), along with the residual on the absolute force magnitude ($\mathbf{{F}_\textrm{res, tot}}$), for increasing pole orders up to 10. The $\mu$ and $\sigma$ remain consistent at the 0.01\% level across increasing pole orders, suggesting negligible improvement in the residual distribution. Given the computational complexity of higher-order terms in the azimuthal harmonic expansion and the lack of significant enhancements in the particle-by-particle forces, we adhere to maintaining $(\ell, m)_\mathrm{max} = 4$ for our TEMP model. This decision is also supported by previous findings, as illustrated in Fig.~7 of \citet{sanders2020models}, where only minor improvement (about $\sim 0.1\%$) in orbit reconstruction is observed for pole orders greater than 4.

\subsubsection{Reconstructions under imposed model symmetry} \label{sec:force_sym}

\begin{table}
\caption{Symmetry types and their implications}
\centering
\begin{tabular}{ccc}
\hline
\textbf{Symmetry} & \textbf{Notation} & \textbf{Non-zero harmonics} \\
\hline
No Symmetry & n & All $\ell$ \& All $m$ \\
Triaxial & t & Even $\ell$ \& Even $m$ \\
Axisymmetric & a & Even $\ell$ \& $m=0$ \\
\hline
\end{tabular}\\

\raggedright
\textbf{Note.} The notation is used as a shorthand to specify the symmetry condition in Fig.~\ref{fig:Fstd_model_violin} and \ref{fig:Fstd_all_violin}. The Non-zero harmonics column lists the spherical harmonic and azimuthal harmonic pole orders that are non-zero for the given symmetry condition.
\label{tab:symmetry_implications}
\end{table}

\begin{figure*}
\centering
\includegraphics[width=\textwidth,]{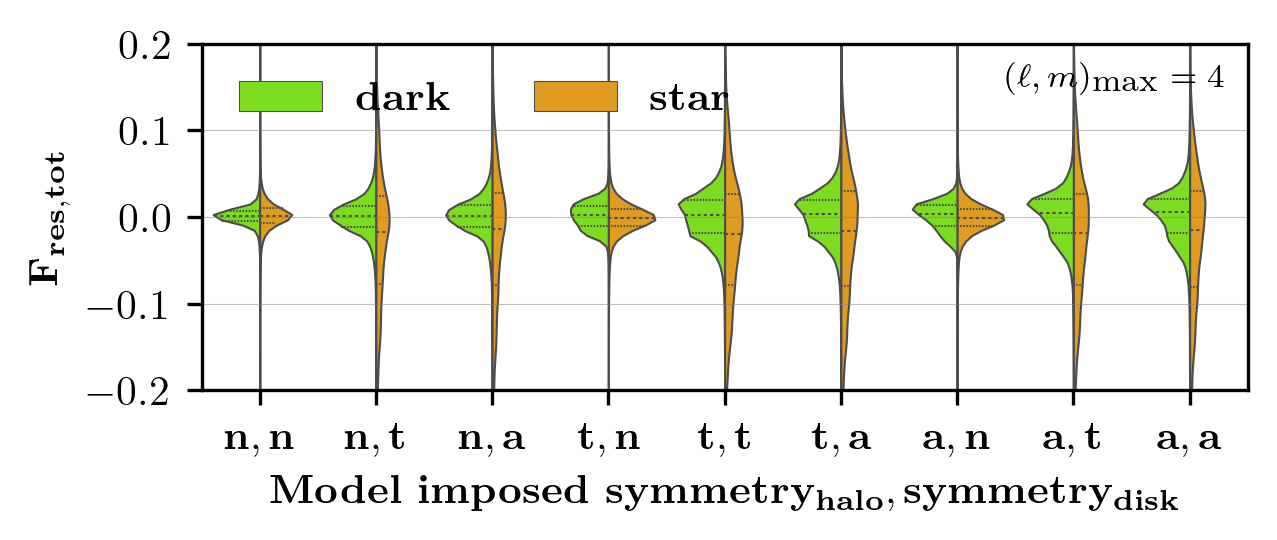}
\caption{{\bf Variation in performance with different symmetry assumptions.} Force residual or the fractional variation from the true force distributions of the absolute magnitude of force for DM particles (green) and star particles (orange) within 50 kpc of the galactic center at the present day. The forces are reconstructed using fixed pole order $(\ell,m)_{\mathrm{max}}=4$ and different symmetry conditions for the halo and the disk. The corresponding symmetry names and notation are provided in Table~\ref{tab:symmetry_implications}. The horizontal dotted lines represent the 25th and 75th quartiles of the distributions, while the dashed line represents the 50th quartile. \label{fig:Fstd_model_violin}}
\end{figure*}

We explore the impact of imposing different symmetry conditions on our TEMP model, with $(\ell, m)_\mathrm{max} = 4$ on the reconstructed force for both the DM and star particles. Such symmetry assumptions are widely used to fit parameterized potential models to the observational data. Additionally, imposing certain symmetries allows us to compute crucial integrals of motion, such as actions \citep{binney2008dynamics}. These symmetry conditions are imposed by setting certain coefficients in the spherical ($\ell, m$) and azimuthal harmonic ($m$) expansions to zero, effectively reducing the number of terms in the model. Table~\ref{tab:symmetry_implications} specifies the poles that remain non-zero under the imposed symmetries. We explore three different symmetries for both the halo and the disk: no symmetry ($n$), axisymmetry ($a$), and triaxial symmetry ($t$), excluding the trivial spherical symmetry for the halo. 

Exploring all possible combinations of imposed symmetries across the halo and disk results in a total of nine different setups. We perform force reconstruction for all the DM particles and stars within the parent halo within 50 kpc of the galactic center at present day. The DM particles are primarily located in the halo and stars are predominant in the disk. The residual distributions obtained from these reconstructions serve as a proxy for the adequacy of our halo and disk models, demonstrating how specific symmetries impact the overall model fidelity.

The Violin plots in Fig.~\ref{fig:Fstd_model_violin} show the distributions of the residual between reconstructed and true force magnitudes for both DM (green) and stars (orange) for the nine models. Each model is denoted by a combination of symmetry conditions on the halo and the disk, as indicated on the x-axis in the format (imposed halo symmetry, imposed disk symmetry), with reference to Table~\ref{tab:symmetry_implications}. The dotted lines represent the 25th and 75th quartiles of the distributions, while the dashed line represents the 50th quartile.

Each violin plot details error distributions, where the majority of the particles consistently exhibit errors within 10\% across all models, with the best performances typically observed in the no symmetry model setup. Specifically, assuming no symmetry on the halo model with any symmetry assumption on the disk, or conversely, no symmetry on the disk with any constraints on the halo model, yields satisfactory results, with most residuals within 2\% for both DM and star forces, respectively. Imposing symmetry conditions generally increases the $\sigma$ and $\mu$ of the reconstructed forces, which vary depending on the specific symmetry applied and the particle type. Symmetry conditions on the halo, such as axisymmetry or triaxiality, for DM particles continues to produce satisfactory results, with errors generally remaining within the 5\% range. Intriguingly, ensuring accurate modeling of the disk without imposing any symmetry constraints often leads to improved accuracy in the reconstructed forces even within the halo region. However, imposing symmetries on the disk leads to higher errors in reconstructed forces, particularly exaggerating the tails of these distributions and higher occurrence of errors exceeding 10\%. In Appendix~\ref{app:force_res_axis} (see Fig.~\ref{fig:Fstd_all_violin}), we present the residual distributions of forces across each Cartesian axis, exhibiting similar trends as observed in Fig.~\ref{fig:Fstd_model_violin}. 
   
In summary, our findings highlight that for modeling halo orbits, strong constraints can be imposed on the disk potential while still achieving accurate orbital reconstructions, even under tight symmetry conditions like axisymmetry. In contrast, for disk orbits, it is imperative to accurately model the disk while employing simpler models for the halo could do sufficiently well. Furthermore, symmetry assumptions could introduce significant biases, particularly when computing integrals of motion for disk orbits. 

\section{Selection of stars and orbit integration} \label{sec:integ_sample}

\begin{figure*}
\includegraphics[width=\linewidth]{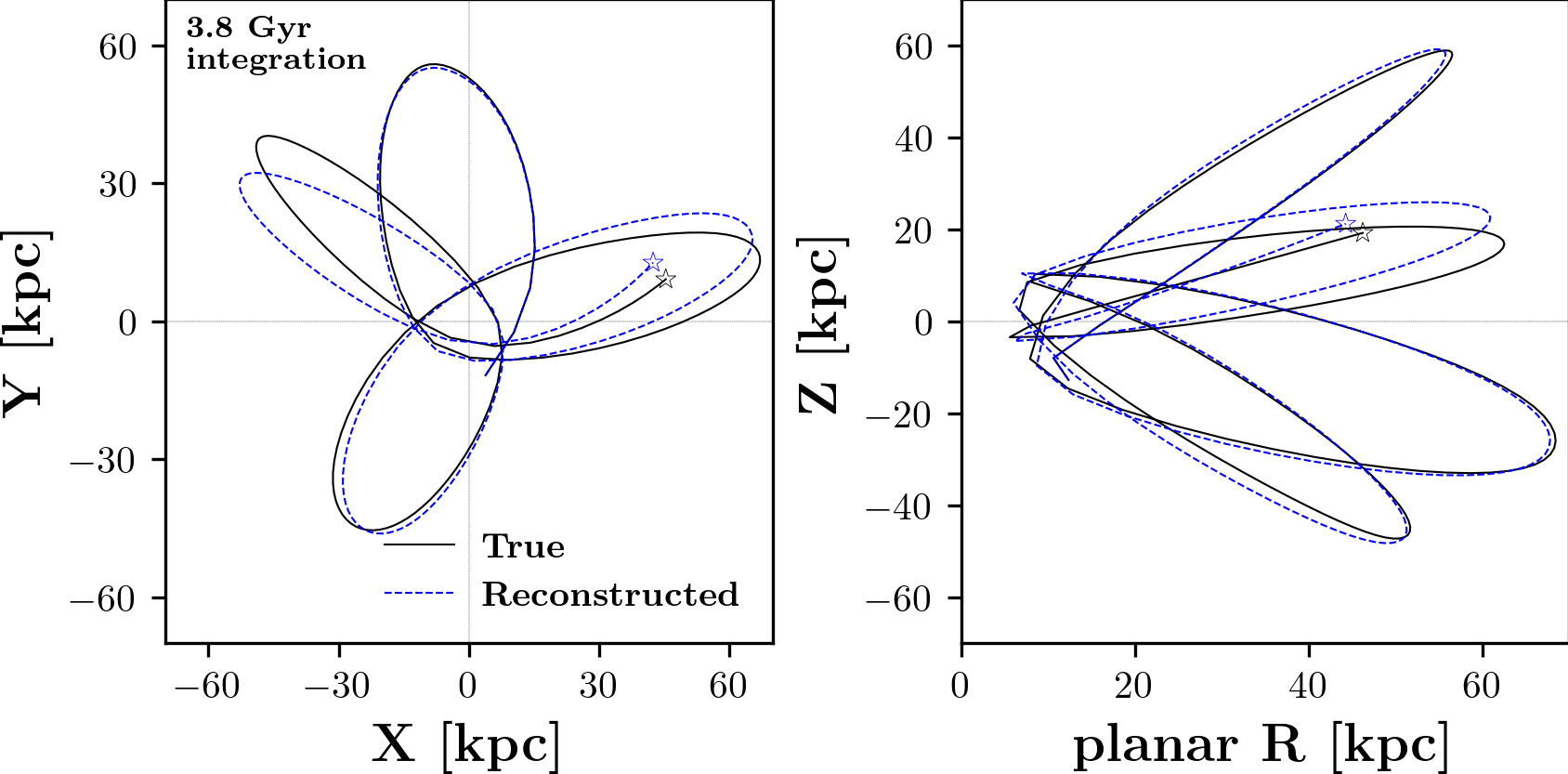}
\caption{{\bf Example of orbit reconstruction.} Real (black solid curve) and reconstructed (blue dashed curve) trajectories for a randomly selected star particle extracted from the simulation and integrated in the time-evolving multipole model from Sec.~\ref{sec:pot_models} for approximately 3.8 Gyr up to the present day in the XY (left) and RZ (right) plane. The trajectories exhibit a close match, indicating a strong agreement between the real and reconstructed paths; positions at present day (i.e. the final timestep of integration) are marked with a star. Fig.~\ref{fig:sample_trajs}, below, shows a sample of 5 other randomly selected orbits with varying orbital periods. All the trajectories are shown in physical coordinates. \label{fig:samp_orbit}}
\end{figure*}

{To explore a representative selection of halo orbits, we select star particles in the simulation that meet the following criteria for tracked halo-like orbits:}  

\begin{itemize}
    \item not associated with any halo except the MW at the selection time ($T \approx 9$ Gyr) and have orbital periods less than 3 Gyr.
    \item formed at least 30 kpc away from the galactic center, i.e not in the MW disk.
    \item have galactocentric distances never exceeding 200 kpc, and within 100 kpc at the present day.
\end{itemize}

{Out of approximately 10 million stars in the simulation, about 3000 fulfill these criteria.} These selected stars are then integrated forward in time for approximately 4 Gyr in the TEMP model with no imposed symmetries, described in Sec~\ref{sec:pot_models}. {The choice of a 4 Gyr integration period is arbitrary in absolute terms, but it ensures that the orbits undergo sufficient dynamical evolution, capturing between 1 and 100 orbital periods.}

To identify stars not associated with any subhalos, we use the {\fontfamily{qcr}\selectfont ROCKSTAR} halo finder \citep{behroozi2012rockstar} to identify DM subhalos and assign stars associated to each subhalo \citep{samuel2020profile, wetzel2023public}. We note that our star selection process may include a few stars (approximately $\sim5\%$) showing indications of being bound to dwarf satellites and/or stream progenitors based on their phase-space distribution and angular momentum along the Z axis. This is a function of the tolerances chosen for determining boundedness of stars to subhalos; additionally particles can be energetically unbound but still associated with subhalos.  

Despite this, we choose to retain these stars in our analysis, hereby referred to as \emph{bound stars}, acknowledging a similar challenge encountered in observational data where determining the gravitational binding of an object can be difficult. These bound stars are highlighted in gray whenever shown, and any reported statistics in this study exclude their contributions.   

We employ a leapfrog algorithm \citep{van1988leap} to integrate sample orbits over the last 4 Gyr of the simulation using our time-dependent potential model. {The integration follows eq.~\ref{eq:force} where the terms on the right-hand side represent:}

\begin{itemize}
    \item \textbf{Gravitational force:} {The first term, $-\vec{\nabla}_{\vec{r}} \Phi(\vec{r},t)$, represents the gravitational force,} which is calculated from the time-dependent potential model $\Phi(\vec{r},t)$. The potential models are saved at discrete time points, and at each integration step, we compute the force experienced by a particle using linear interpolation between adjacent snapshots of the potential.

    \item \textbf{Cosmological expansion:} {The second term, $-\frac{\dot{a} (t)}{a (t)} \vpec$, accounts for the effect of cosmological expansion, where $a(t)$ is the scale factor and $\dot{a}(t)$ is its time derivative (the Hubble parameter). We compute this term by evaluating the time-dependent scale factor $a(t)$ at each integration step, applying a velocity ``kick'' corresponding to the cosmological effect on peculiar velocities.}

    \item \textbf{Motion of the center-of-mass:} {The third term, $-\frac{d \vec{u}_\textrm{COM}}{dt}$, captures the acceleration due to the motion of the galaxy's center-of-mass (COM). To compute this term, we use spline fits to the time series of the COM velocity, $\vec{u}_\textrm{COM}$, for the galaxy at each snapshot. These spline fits provide a smooth representation of the COM motion, allowing us to calculate the velocity change (kick) imparted to each particle at every time step.}
\end{itemize}

{For each particle,} the leapfrog integration carefully chooses a timestep ($\Delta t$) to resolve the pericenter of the orbit. The $\Delta t$ is computed based on the initial conditions of the orbit using the present-day potential to estimate the pericenter, ensuring sufficient resolution near critical points as:

\begin{equation} \Delta t = \frac{\text{tolerance} \times v_p}{a_p} \end{equation}

{where $v_p$ is the velocity at pericenter and $a_p$ is the acceleration at pericenter, both derived from the potential. This guarantees that the pericenter is well-resolved, preventing integration errors from accumulating in close encounters. The fixed timestep ensures that the integration errors from the leapfrog method are small compared to the force approximation errors introduced by the potential model. As a symplectic integrator, leapfrog bounds errors over time, making it well-suited for long-term orbit stability \citep{yoshida1990construction, preto1999class}. In contrast, force errors from the potential model accumulate and dominate the overall inaccuracy.}

Fig.~\ref{fig:samp_orbit} plots the real (solid black) and reconstructed (dashed blue) trajectory of a {randomly} selected star from {our sample} integrated for approximately 3.8 Gyr to the present day, in the XY (left) and RZ (right) plane. Both the trajectories start from the same point and the final positions at present day are marked with a star. Overall, the orbit exhibits a close match, indicating a strong agreement between the real and reconstructed paths. 

\section{Accuracy of Reconstructed Orbits} \label{sec:recon_orbits}
 
 In this section, we quantify the accuracy of reconstructed orbits for the sample of $\sim3000$ stars described in Sec.~\ref{sec:integ_sample} with our TEMP model consisting of a spherical harmonic expansion for the DM halo and azimuthal expansion for the galactic disk (see Sec.~\ref{sec:pot_models}). We quantify the spatial and temporal dependence of relative error between true and reconstructed positions for each star (Sec.~\ref{sec:orbit_error_metric}). We also compare integrals of motion such as the total energy and angular momentum (Sec.~\ref{sec:integ_motion}), which are robust proxies for quality of our reconstructed orbits and statistically quantify ``failure'' modes based on a 100\% error in recovering total angular momentum for an orbit.   

 \subsection{Relative position error metric} \label{sec:orbit_error_metric}

\begin{figure*}[h]
    \centering
    \includegraphics[width=\textwidth]{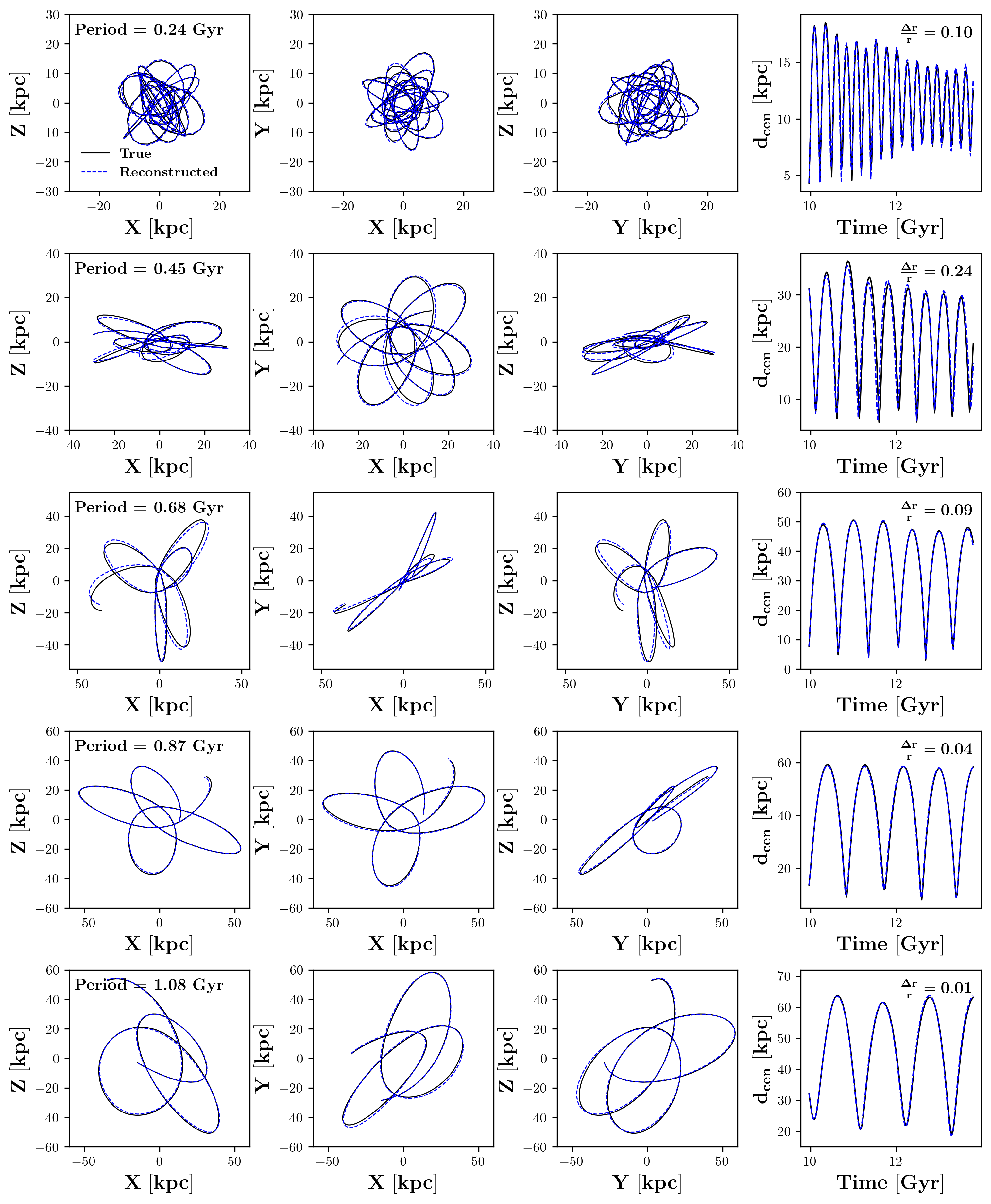}
    \caption{{\bf Performance over a range of orbital periods.} Randomly selected reconstructed orbits (blue dashed lines) compared with the original trajectories (black solid lines) of stars from the simulation across Cartesian planes (first three columns from left) in physical coordinates, and in terms of the distance from the host center over time (rightmost column). The orbits are plotted in order of increasing orbital period (listed in the top left corner in the leftmost panel in each row). The relative position error (see eq.~\ref{eq:error_metric}) at the final time step is listed in the top right corner of the rightmost plot in each row. The method reconstructs orbits accurately across a factor of 20 in orbital periods and 100 in galactocentric distance.}\label{fig:sample_trajs}
\end{figure*}

We evaluate the TEMP model's performance in reconstructing orbits by measuring the {relative position error} for each orbit trajectory at time $t$ as

\begin{equation} \label{eq:error_metric}
    \frac{\Delta r}{r}(t) = \frac{||\vec{r}_\textrm{reconstructed}(t) - \vec{r}_\textrm{true}(t)||}{||\vec{r}_\textrm{true}(t)||}.
\end{equation}

Here, $\vec{r}_\textrm{true}(t)$ and $\vec{r}_\textrm{reconstructed}(t)$ represent the true and reconstructed 3D positions of each particle at time $t$ in physical coordinates. This metric quantifies trajectory errors for each particle, explicitly considering orbit phase. While \cite{sanders2020models} employ a similar metric, they base it on the error relative to the \textit{time-averaged} radius of the orbit. This measure tends to underestimate errors at pericenter, where a small phase error can lead to a large position or velocity error, and overestimate them at apocenter, where a large error in phase maps to a much smaller position or velocity error. Our metric directly compares reconstructed and true positions at each time step, offering a more accurate evaluation of trajectory fidelity throughout the orbit. We report this metric at the final time step as $\frac{\Delta r}{r}$. Additionally, we normalize times based on the orbit period, computed using a fast Fourier transform of the particle's true trajectory to identify the dominant frequency.     

Fig.~\ref{fig:sample_trajs} shows a selection of 5 other randomly chosen orbits, arranged by increasing orbital period. The trajectories are presented in Cartesian planes across the first three columns, with the final column illustrating the distance from the center as a function of time. Upon visual examination, the majority of orbits are reproduced with a reasonable level of accuracy. However, orbits with multiple orbital periods and an average orbital distance within 20 kpc exhibit higher errors, exceeding 10\%. Additionally, deviations in other trajectories may result from interactions with substructures that are not adequately resolved with our low order expansion $(\ell, m)_{\textrm{max}} = 4$. Subhalos on the order of a few kiloparsecs in size require a significantly higher harmonic order \citep{lowing2011halo}.

\subsubsection{Phase-space dependence at the final timestep}

\begin{figure*}
\includegraphics[width=\textwidth,]{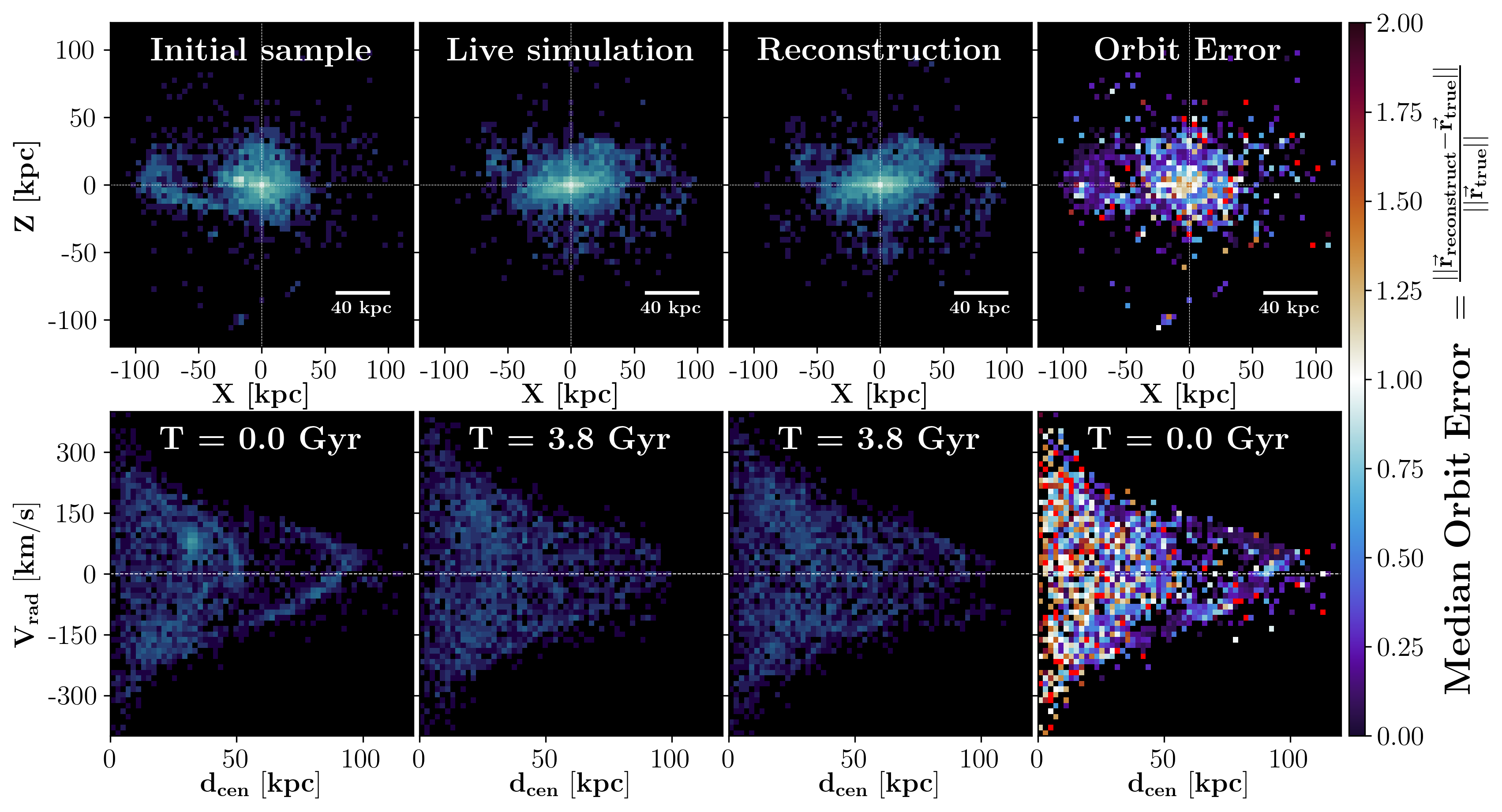}
\caption{{\bf Preservation of phase-space distribution.} Initial sample of all the selected stars (leftmost column), their distribution in the live simulation at ``present day'' (second from left), and the reconstructed distribution using the TEMP model for orbit integration (third from left). Distributions are shown in the XZ plane (top row) and for galactocentric distance vs.radial velocity (bottom row). The rightmost column shows the median value of error metric for the initial distribution of sample stars in each bin. Most orbits have errors below 15\%, with no angular dependence in the error metric. Errors are larger for particles closer to the center, deep in the disk potential. \label{fig:phase_space_error}}
\end{figure*}

Fig.~\ref{fig:phase_space_error} illustrates the phase-space distribution of selected stars for integration (column 1), depicting their distributions in the XZ plane (row 1) and the total distance from the center and radial velocity plane (row 2) in the live simulation (column 2) and their reconstructed distributions (column 3) using the TEMP model after 3.8 Gyr of orbit integration. The last column exhibits the spatial dependence in the error metric, representing the median value of the relative orbit error ($\frac{\Delta r}{r}$) defined by eq.~\ref{eq:error_metric} at the final time step for the stars in the initial sample.  

In general, the orbit reconstruction demonstrates consistency, with errors typically below 15\% after 3.8 Gyr of integration, and no significant angular dependence at different distances in the error metric. However, there is a systematic issue with reconstruction accuracy in the inner regions of the galaxy (within 15 kpc), likely due to the stronger influence of the disk, which is consistent with the errors in reconstructed forces in these regions (Fig.~\ref{fig:Fstd_n4n4}). Additionally, no significant trends are observed with radial velocity.

We also note the presence of stars from a small satellite galaxy in our sample, approximately 40 kpc away with a radial velocity of 100 km $\textrm{s}^{-1}$. This satellite, while not associated with any subhalos in the halo catalog, appears to have bound stars in the sample. Our sample also contains stars from an unbound stellar stream with a small bound progenitor is identified in our sample (see Fig.~\ref{fig:phase_space_error}). As the simulation progresses, the satellite undergoes tidal disruption and phase mixing, resulting in a median orbit error in our reconstruction of roughly order 1, attributed to the TEMP model's ignorance of self-gravity in the system. However, high-fidelity orbit reconstruction is observed for streams with errors below 5\% along the stream track, except for the bound progenitor, which again exhibits an error of order 1. This emphasizes the critical role of self-gravity and the stripping times of stars. Ignoring these factors results in biased orbits. 

\subsubsection{Positional errors at the final timestep}

\begin{figure}
\includegraphics[width=\linewidth]{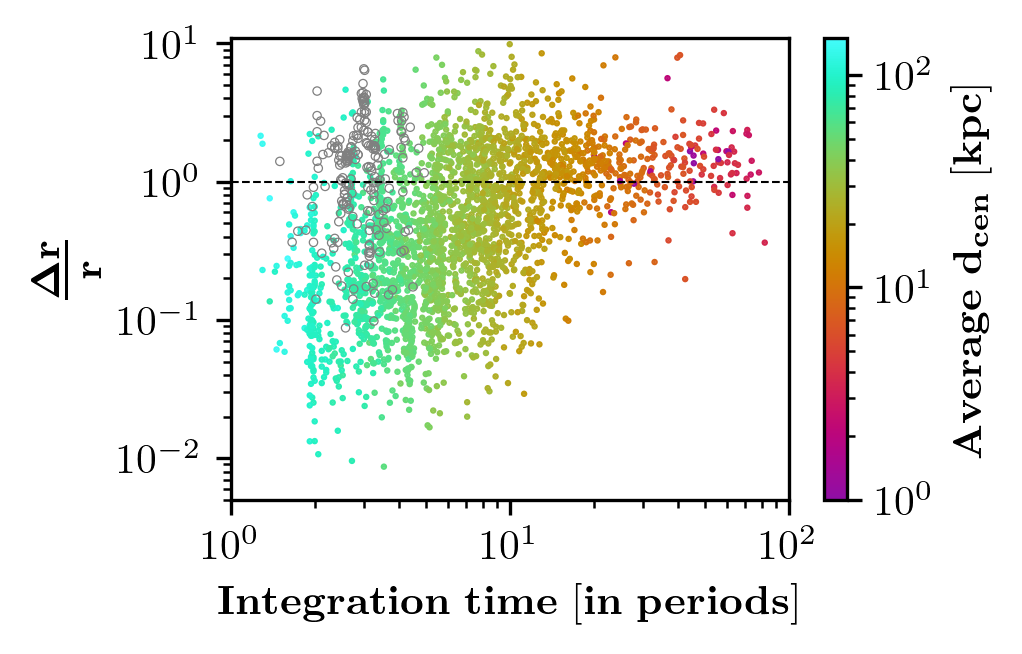}
\caption{{\bf Performance over multiple orbital periods.} Relative position error at the final time step of integration (approximately 3.8 Gyr) versus the total integration time in terms of the number of periods passed for each orbit. Orbits are color-coded by the average distance from the center for each star. About 70\% of orbits have relative error less than one at the end of the integration time. Most orbits with $\frac{\Delta r}{r} \geq 1$ are integrated over $\geq 10$ periods and are located close to the center of the host. The bound stars identified from Fig.~\ref{fig:phase_space_error}, which are expected to have larger error due to their unmodeled progenitor galaxy, are shown with gray markers.\label{fig:delta_r_final}}
\end{figure}

Fig.~\ref{fig:delta_r_final} plots $\frac{\Delta r}{r}$ at the final time step of integration as a function of integration time in periods passed for each orbit, color-coded by the true average distance from the host during the integration time. Also, the bound stars are shown with gray markers and majority of them show large errors. Approximately 70\% of orbits retain phase-space information after the integration time, with $\frac{\Delta r}{r} \leq 1$ . Comparing exact errors, only 13\% of the recovered orbits have errors less than 10\% (see Table.~\ref{tab:error_integ_motion} for more statistics). 

{Most of the orbits with higher relative errors ($\frac{\Delta r}{r} \geq 1$) are integrated for many dynamical times (over 10 periods) and are situated close to the center of the host (within 15 kpc). This is consistent with the fact that the largest fractional errors are in the central region (see Fig.~\ref{fig:Fstd_n4n4}). Additionally, these orbits frequently cross regions with stronger gravitational gradients, which can exacerbate the accumulation of integration errors.}

{Some stars on orbits with larger average galactocentric distances ($\sim100$ kpc), also have larger errors. There are a few reasons for this. First, the orbit maybe highly eccentric, in which case the average distance may not accurately represent the orbit’s proximity to the center over time. Second, the phase sensitivity of the error metric amplifies errors near pericenter, leading to higher $\frac{\Delta r}{r}$ values, also matters most for highly radial orbits. Lastly, the omission of smaller-scale structures, such as subhalos, in our model may contribute to the discrepancies observed, as these can cause additional perturbations to the star particle in the simulation not accounted for in the TEMP model.}

\subsubsection{Temporal dependence in the error metric}

\begin{figure*}
\includegraphics[width=\textwidth]{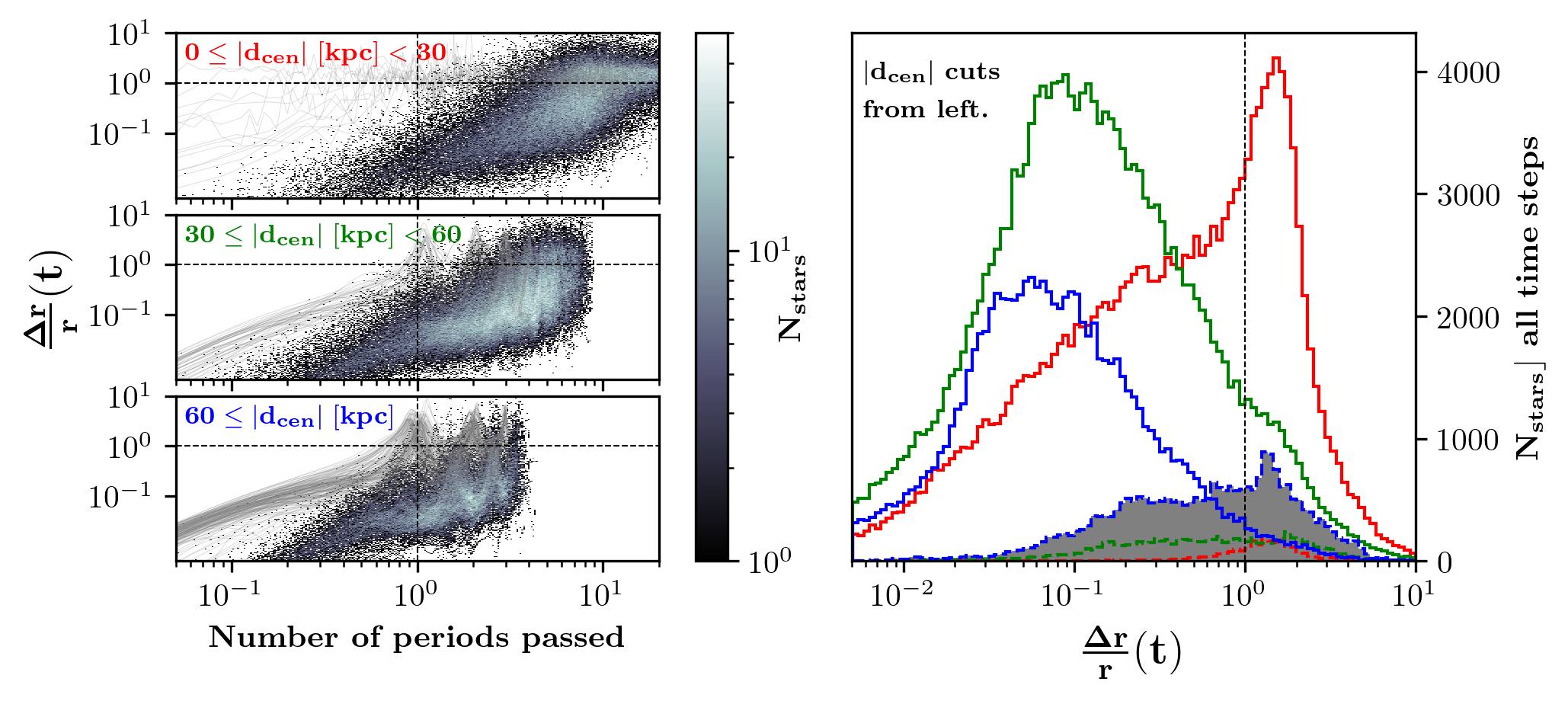}
\caption{{\bf Performance of orbit reconstruction over time at different average distances.} {\it Left:} Temporal evolution of relative position error as a function of number of periods passed for each orbit, divided in rows based on the average orbital distance from the center. Distance ranges are indicated in the top left corner of each row. {\it Right:} Distribution of relative position error for all orbits at each time step, color-coded by the distance ranges shown on the left.  Most errors are within 10\% for 2--3 periods across all distance cuts, while orbits with $\frac{\Delta r}{r} (t) \geq 1$ predominantly occur for orbits closer to the center after 5--6 periods. Additionally, the error trajectories oscillate with orbital phase by nearly an order of magnitude, as is most clearly visible for the longest-period orbits in the bottom left panel: $\frac{\Delta r}{r} (t)$ is higher at pericenter and lower at apocenter. Error trajectories and distributions for bound stars are marked with gray lines (left) and a gray histogram (right). \label{fig:delta_r_evolv_hist}}
\end{figure*}

In addition to the error metric at the final time step, it's important to compare how the metric evolves over time to test the stability and reliability of our reconstruction approach at different time steps. This temporal perspective allows us to evaluate the long-term behavior of orbits and identify any potential sources of bias or inaccuracies that may arise over extended periods of integration. 

Fig.~\ref{fig:delta_r_evolv_hist} depicts the temporal evolution of relative position error ($\frac{\Delta r}{r} (t)$) as a function of total number of periods passed for each orbit on the left, organized into rows based on average orbital distance from the center. The top row includes orbits between 0-30 kpc, the middle row 30-60 kpc, and the bottom row $\geq 60$ kpc.  The histograms on the right show the $\frac{\Delta r}{r} (t)$ for all orbits at each time step, color-coded by the distance cuts. The error metric trajectories and distributions for bound stars are also shown in gray.  

Most of the orbital errors (approximately 90\%) remain within 10\% for up to 2 periods across all distance cuts, indicating consistent performance of our method in capturing orbital dynamics. Notably, instances where $\frac{\Delta r}{r} (t) \geq 1$ predominantly occur for stars closer to the center after 5-6 periods have passed. Additionally, about 80\% of the outer orbits (average orbital distance $\geq 30$ kpc) exhibit $\frac{\Delta r}{r} (t) \leq 0.1$. Other significant errors are observed in orbits associated with the bound satellite and stream progenitor (marked with gray lines) identified in the sample (refer to Fig.~\ref{fig:phase_space_error}).

Interestingly a notable trend emerges wherein orbits approaching their pericentric passage exhibit higher $\frac{\Delta r}{r} (t)$, whereas lower errors are observed at apocenter. This trend is particularly prominent in the outermost halo ($\geq 60$ kpc), where the $\frac{\Delta r}{r} (t)$ trajectory predominantly displays a triangular wave pattern. The error peaks near pericenter and valleys at apocenter due to the smallest division factor in our metric at pericenter, amplifying the $\frac{\Delta r}{r} (t)$. Additionally, the same phase error in the orbital plane magnifies the positional error into a larger discrepancy at pericenter compared to apocenter.

Moreover, the faster tangential velocity at pericenter contributes to increased velocity errors, while slower tangential velocities at apocenter allow the phase to synchronize between reconstruction and true trajectories. \citet{sanders2020models, d2022uncertainties,santistevan2023orbital} also observed similar trends, with increased errors in pericenter reconstruction of satellite orbits, where they compared the true and reconstructed pericenter position neglecting the phase error. 

While higher positional errors occur at pericentric passage, in a summary statistic, this discrepancy may not be significant. Most orbits spend more time near apocenter, so for an arbitrary selected final time step, most stars will have lower positional errors. However, this discrepancy would bias positional errors if one is examining only resonant orbits that reach pericenter at the same time.

\subsubsection{Pericenter and apocenter comparison}

\begin{figure*}
\includegraphics[width=\textwidth]{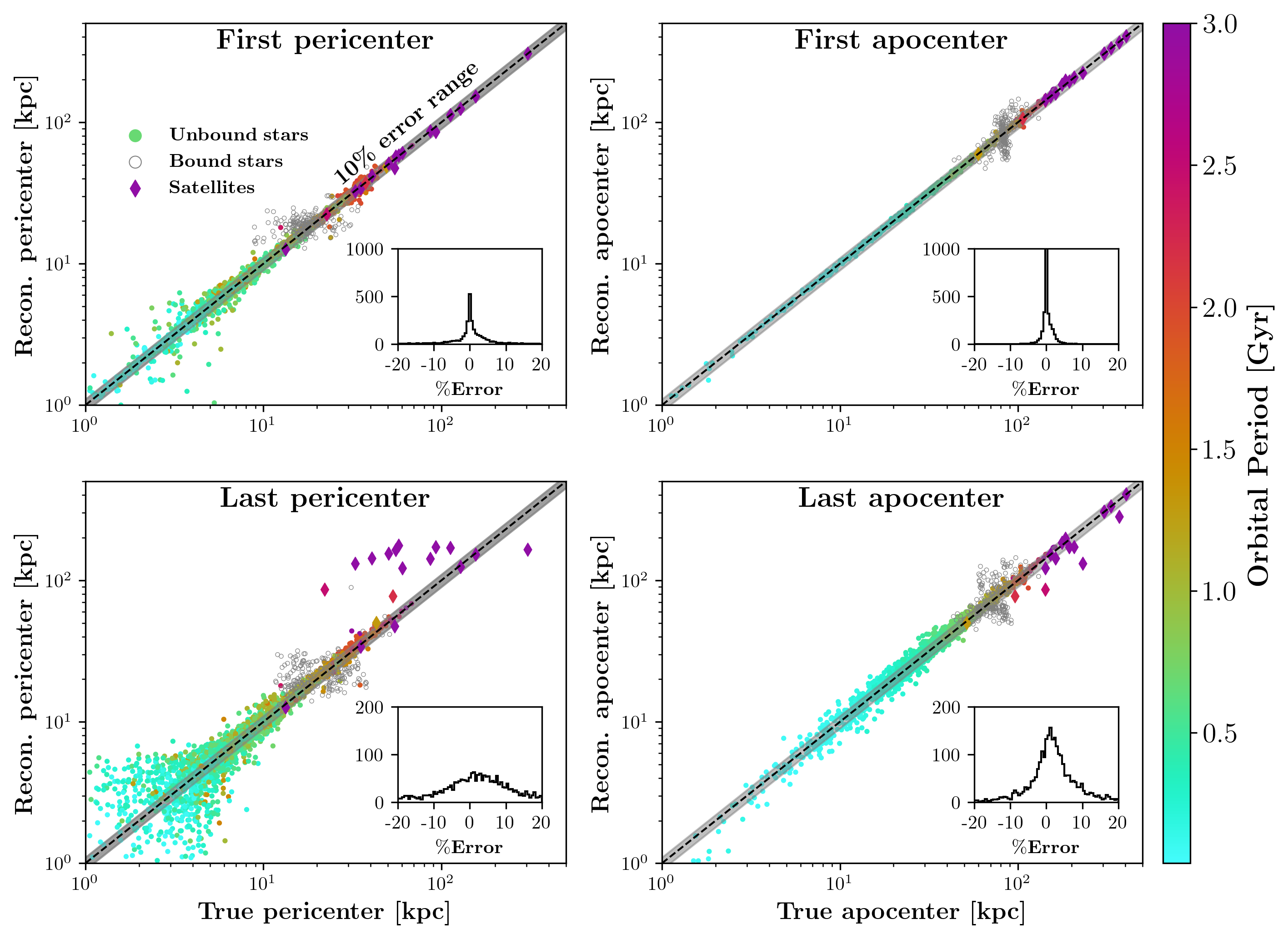}
\caption{{\bf Reconstruction of pericenter and apocenter distances.} Reconstructed versus true pericenter distances (left column) and apocenter distances (right column) for the first (top row) and last (bottom row) passages, respectively. Orbits of unbound stars (solid circles) and luminous satellites (diamonds) are color-coded by their orbital period. Most orbits exhibit errors within 10\% (shaded gray region around dashed 1:1 line) at the first passage, with higher errors observed closer to the center. Notable discrepancies in the pericenter and apocenter distances, such as those observed at 20 kpc and 80 kpc, respectively, primarily originate from the bound parts of the stream included in the sample (gray open circles; see also Fig.~\ref{fig:phase_space_error}). By the final passage, pericenter distances in the inner regions have more scatter than apocenter distances, although errors for orbits with pericenter beyond 30 kpc remain within 10\%. The insets in each panel plot the distribution of percentage errors between reconstructed and true values for unbound stars. The reconstructed satellite orbits overestimate the last pericenter due to the absence of a dynamical friction prescription in our model. \label{fig:peri_apo}}
\end{figure*}

\citet{santistevan2023orbital, santistevan2024modelling} used the MW-mass galaxies from the \textit{Latte} suite of baryonic-cosmological simulations \citep{wetzel2023public} and beginning at present day, backward integrated the center-of-mass (COM) positions and velocities of luminous satellites orbiting the main host. They employed a static MW-mass potential and ignored dynamical friction. They showed that recovering the first pericentric and apocentric distance through orbit reconstruction has a 20-40\% uncertainty with higher uncertainties in pericentric distance.

Similarly, \citet{d2022uncertainties} used a controlled MW host with no massive mergers from the Elvis suite of DM only simulations \citep{garrison2014elvis}, backward integrated subhalo COM positions and velocities accounting for the mass growth of the main halo while keeping the shape of the potential fixed. They also used a prescription for the dynamical friction experienced by the satellites. They reported the fraction of satellite orbits that have less than 30\% error in the pericentric and apocentric distances. They found that 70\% of the satellites were below this threshold for their first pericentric distances, and only 55\% of subsequent pericentric distances less than 30\% error, while 90\% of first apocenters and 75\% of subsequent apocenters had similar errors below 30\%.

Motivated by these findings, we backtrack the COM positions and velocities of luminous satellites ($\textrm{M}_\star > 0$ \Msol{}) at that are within the virial radius of the main halo and have a total mass less than $10^{10}$ \Msol{} the present day to 4 Gyr ago. We then forward integrate to the present day in our TEMP model, ignoring dynamical friction (similar to \citet{santistevan2024modelling}). Many of these satellites can be much more massive 4 Gyr ago.

Fig.~\ref{fig:peri_apo} compares the reconstructed and true pericenter distances (left column) and apocenter distances (right column) for the first (top row) and last (bottom row) pericentric passages, respectively. Unbound star (solid circles) and luminous satellite (diamonds) orbits are color-coded based on their orbital period, with gray markers indicating bound stars within our selection sample. The inset in each panel shows the distribution of percentage error in each reconstructed property. At first passage, most orbits both in pericenter and apocenter closely align with the 1-1 line, exhibiting errors within 10\% (insets in top row). However, typically higher errors are noticeable nearer to the galactic center, along with errors originating from the satellite and stream progenitor in both pericenter and apocenter distances (gray scatter points), as highlighted in Fig.~\ref{fig:phase_space_error}, with notable discrepancies particularly evident at pericenter and apocenter distances around 20 kpc and 80 kpc, respectively.

As the simulation progresses, pericenter distances exhibit greater variability compared to apocenter distances by the final passage (insets in bottom row). Nevertheless, errors for orbits with pericenters beyond 30 kpc remain within the 10\% threshold. Similar errors in pericenter and apocenter distances are noted in \citet{sanders2020models}. Accurate reconstruction of pericenter and apocenter distances is crucial, as they are fundamental properties widely used in studying small scale structure formation and disruption within a galaxy \citep[e.g.][]{barber2014orbital,simon2018gaia, shipp2023streams}.

Luminous satellite orbits follow similar trends to unbound stars. We recover 85\% and 98\% of pericentric and apocentric distances for luminous satellites to errors within 10\%. These recovery rates are overall better than static potential models that only account for the mass growth. Subsequent pericenters are harder to recover, but we still achieve 70\% recovery of subsequent apocenters within 10\%. The reconstructed satellite orbits overestimate the last pericenter due to the absence of a dynamical friction prescription in our model.

In summary, our reconstruction method demonstrates overall success, maintaining errors below 15\% after 3.8 Gyr of integration, with approximately 70\% of orbits exhibiting a relative error below 1. However, accuracy issues arise within the inner galaxy regions, likely due to the disk's stronger influence and challenges in reconstructing orbits of bound substructures. Notably, the stable, quiescent merger history in this simulation ensures that the potential undergoes minimal non-adiabatic changes, contributing to the robustness of the reconstruction process. The temporal evolution analysis reveals that most orbital errors remain within 10\% for up to 2 periods across all regions, but deteriorate after 5-6 periods for orbits within 30 kpc. We notably observe higher positional errors at pericentric passages, consistent with prior studies. Our reconstruction method generally accurately reproduces orbits up to the initial pericentric and apocentric passages. However, errors in reconstructed pericenter tend to increase for orbits with pericentric distances closer to the galactic center, while the apocenters are generally recovered more accurately \citep{sanders2020models, d2022uncertainties, santistevan2024modelling}. 

\subsection{Reproducing integrals of motion}\label{sec:integ_motion}

\begin{figure*}
\includegraphics[width=\textwidth]{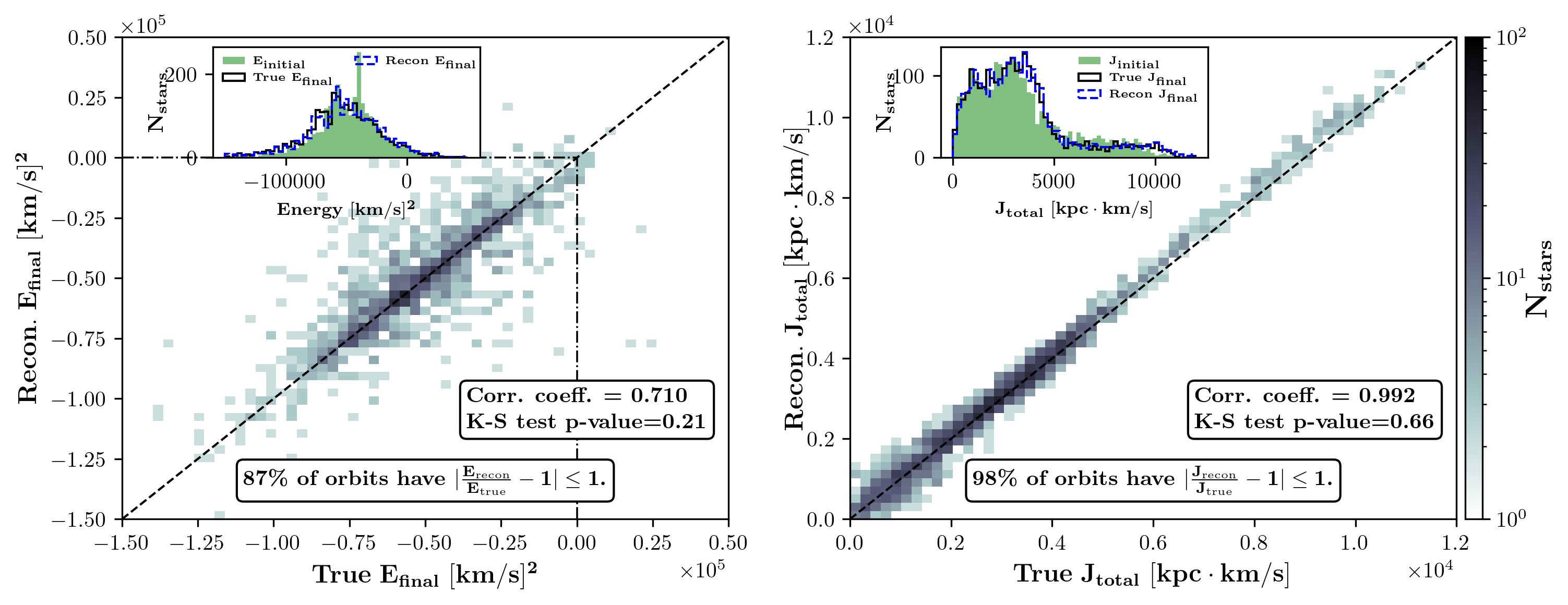}
\caption{{\bf Recovery of integrals of motion.} {\it Left:} Reconstructed final energy versus true final energy. The inset in the top left corner displays the distribution of the initial energy (green), true final energy (black), and reconstructed final energy (blue). The dashed-dotted line indicate the energy threshold ($E_\textrm{tot} = 0$ (km/s)$^{2}$) above which particles are unbound to the main halo. {\it Right:} Similar to the left panel, but for the total angular momentum ($J_\textrm{total}$). $J_\textrm{total}$ exhibits less scatter and tighter correspondence with a stronger 1:1 correlation (black dashed diagonal line) compared to the energy. Approximately 98\% of orbits have final angular momentum errors below 100\%, and 87\% have final energy errors below 100\%. Both of the panels have high correlation coefficients ($\geq 0.7$) and p-value greater than 0.5 for the Kolmogorov-Smirnov test, suggesting no significant differences between either of the reconstructed and true distributions.
\label{fig:energy_mom}}
\end{figure*}

To evaluate the quality of our orbit reconstructions, we analyze approximate integrals of motion: energy and total angular momentum ($J_\textrm{total}$). While energy is not strictly conserved due to the time-dependent potential and tends to grow adiabatically, the total angular momentum remains approximately conserved over time \citep{binney2008dynamics, sanders2020models}. The position-based error, particularly dependent on a particle's radial distance from the halo center, can vary rapidly along an orbit \citep{lowing2011halo}.

Fig.~\ref{fig:energy_mom} plots the comparison between reconstructed and true final energy (left panel) and final total angular momentum ($J_\textrm{total}$) (right panel). The top left insets in both panels show the distributions of initial (green), final from simulation (black), and final from reconstruction (blue) for both energy and $J_\textrm{total}$. Notably, $J_\textrm{total}$ exhibits less scatter and tighter 1-to-1 correspondence with a correlation coefficient of almost 0.99 compared to energy, which has a correlation coefficient 0.71. 

\begin{table}
\centering
\caption{{\bf Accuracy of recovered integrals of motion.} Summary of the fraction of orbits for which the errors in total energy ($E_\textrm{tot}$) and total angular momentum ($J_\textrm{tot}$) are below various thresholds.}
\begin{tabular}{|c|cccc|}
\hline
\multirow{2}{*}{\textbf{Error in:}} & \multicolumn{4}{c|}{\textbf{\begin{tabular}[c]{@{}c@{}}Fraction of orbits recovered \\ with errors under:\end{tabular}}} \\ \cline{2-5} 
                                    & $\leq 10\%$                  & $\leq 25\%$                  & $\leq 50\%$                 & $\leq 100\%$                 \\ \hline
$\mathbf{E_{tot}}$                  & 0.43                         & 0.61                         & 0.75                        & 0.87                         \\ \hline
$\mathbf{J_{tot}}$                  & 0.7                          & 0.85                         & 0.93                        & 0.98                         \\ \hline
$\mathbf{\frac{\Delta r}{r}}$ & 0.13 & 0.33 & 0.5 & 0.7 \\ \hline
\end{tabular}\label{tab:error_integ_motion} \\

\raggedright

\textbf{Note.} Errors are computed after roughly 4 Gyr of integration as the fractional difference between the recovered and true values of each property, defined as $(\text{recovered} - \text{true}) / \text{true} \times 100\%$. The table shows the fraction of orbits where this error is below each specified threshold.
\end{table}

We find that 87\% of orbits have final energy errors below 100\%, with 75\% below 50\%, 61\% below 25\%, and 43\% below 10\% (see Table~\ref{tab:error_integ_motion}). In contrast, total angular momentum ($J_\textrm{total}$) errors are lower, with 98\% of orbits having errors below 100\%, 93\% below 50\%, 85\% below 25\%, and 70\% below 10\%. This indicates that $J_\textrm{total}$ is more robustly conserved, reflecting the relative stability of angular momentum in our dynamic models. The histograms show the presence of bound substructure in the initial energy distribution (over 200 particles in a single bin), which affects the conservation accuracy. 

Our correlation between true and reconstructed final energy is slightly weaker compared to that reported in \citet{sanders2020models}. This disparity can be attributed to the presence of a live baryonic disk actively forming stars, coupled with realistic feedback models, resulting in a larger non-conservation of energy.    

We perform a Kolmogorov-Smirnov (KS) test to assess whether the reconstructed and true distributions of final energy and $J_\textrm{total}$ come from the same distribution. The resulting p-value for energy and $J_\textrm{total}$ is 0.21 and 0.35, suggesting no significant differences between either of the distributions. Additionally, the correlation coefficient between true and reconstructed angular momentum along the Z-axis is 0.995, with a KS test p-value of 0.5, further supporting the consistency between the reconstructed and true distributions.

\subsection{Instantaneous failure based on total angular momentum}

\begin{figure}
\includegraphics[width=\linewidth]{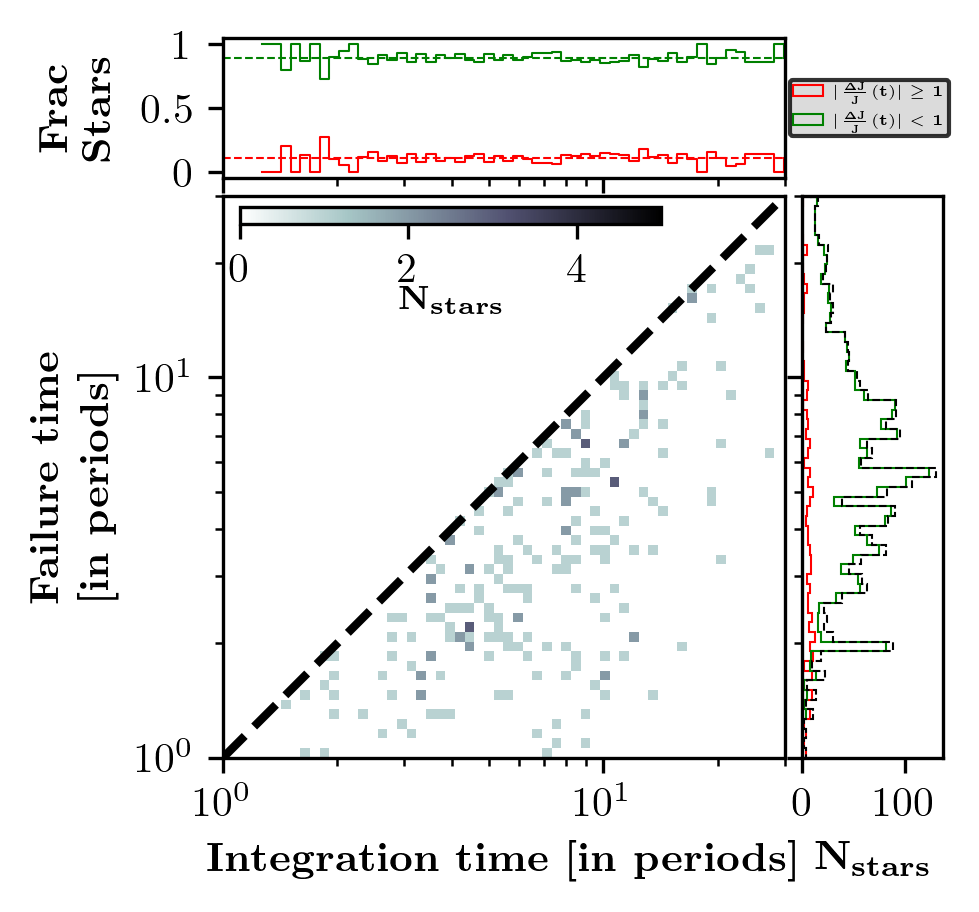}
\caption{{\bf Number of periods before ``failure'' of orbit reconstruction.} Failure time is defined as the moment when the instantaneous error in total angular momentum between reconstructed and true values exceeds 100\%. It is plotted against integration time (in orbital periods), exclusively for orbits that fail (approximately 10\%). The distribution of failure times (red), integration times for orbits that never fail (green), and all the integration times (black) are displayed on the right side. The histogram above the main panel illustrates the fractional integration times for orbits that never failed (green) and those that failed (red), with each bin scaled so that the total across both categories for each bin is 1. Notably, the distribution of failure times appears to be independent of integration time, as indicated by the nearly constant fraction of failures and successes across integration time bins (dashed lines) in the top panel. \label{fig:Tfail}}
\end{figure}

The high fidelity of $J_\textrm{total}$ in our orbit reconstructions, compared to reconstructed positional error (see Fig.~\ref{fig:delta_r_evolv_hist}) and energy (see Fig.~\ref{fig:energy_mom}) motivates establishing a criterion for instantaneous failure in orbit reconstruction. We propose defining failure time as the moment when the instantaneous error in $J_\textrm{total}$ between reconstructed and true values exceeds 100\%. This criterion provides a robust measure to pinpoint instances where orbit recovery becomes unattainable due to the complete loss of phase-space information \citep{arora2022stability}.

Fig.~\ref{fig:Tfail} illustrates failure time plotted against integration time in number of periods for orbits that fail based on the aforementioned criteria, with only approximately 10\% of orbits failing. Surprisingly, the distribution of failure times seems to be independent of integration time. However, one might expect that orbits integrated for longer periods would exhibit a higher likelihood of failure. The top panel illustrates the distribution of integration times for orbits that failed (red) and those that never fail (green), with each bin scaled to represent the fraction of stars. Remarkably, the success rate remains consistently high at around 90\%, while the failure rate remains low at approximately 10\%, regardless of integration time. The side panel plots the distribution of failure times (red), integration times for orbits that never fail (green), and all integration times (black), revealing no discernible pattern in failure time distribution. This observation suggests that neither longer nor shorter integration times significantly affect the likelihood of failure. The independence suggests failure to reconstruct a specific orbit is not due to time dependence of the global potential, but rather to more localized factors like subhalo interactions altering angular momentum or orbits passing through the baryonic disk.

Prolonged integration spanning multiple periods can be effective in preserving angular momentum-dependent properties, such as the shape of orbits, despite the high errors in reconstructed positions. This underscores the robustness of orbit reconstruction techniques in capturing essential dynamical features over extended integration times. 

\subsection{Dependence on the sampling interval} \label{sec:samp_interval}

\begin{figure}
\includegraphics[width=\linewidth]{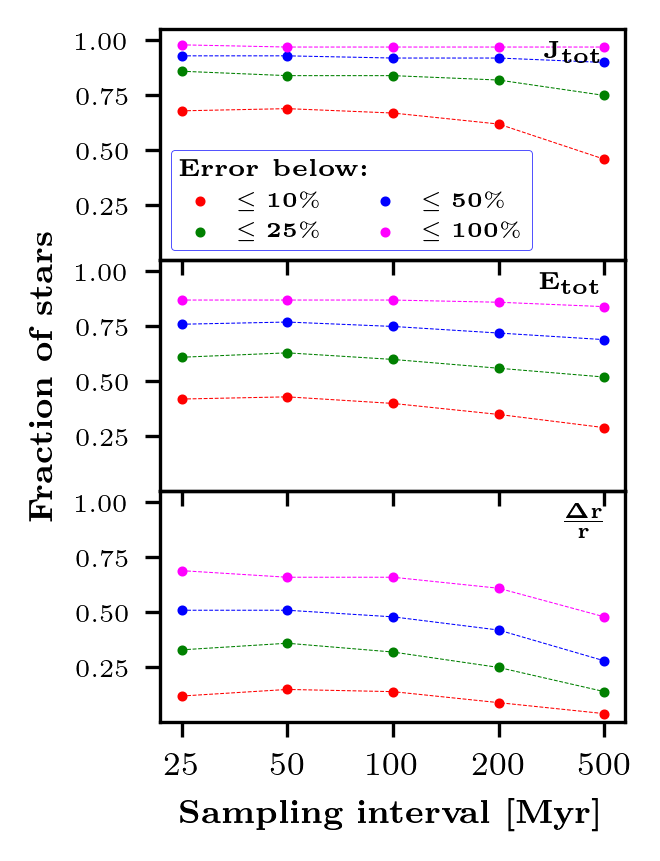}
\caption{{\bf Quality of orbits with lower temporal cadence.} Fraction of orbits with error below the ranges marked with colors in recovered property: total angular momentum $J_{\text{tot}}$ (top row), total energy $E_{\text{tot}}$ (middle row), and positional error metric $\frac{\Delta r}{r}$ (bottom row) as a function of how frequently the potential is sampled for an orbit integration of 4 Gyr. $J_{\text{tot}}$ and $E_{\text{tot}}$ can be accurately recovered with higher sampling intervals (less frequent snapshots), with significant deterioration only observed at a 500 Myr sampling interval. $\frac{\Delta r}{r}$ is more sensitive, showing noticeable decrease at a 100 Myr sampling interval. \label{fig:intervals}}
\end{figure}

The simulation snapshots in FIRE-2 simulations are saved rather frequently—approximately every 25 Myr—while this isn't usually the case for baryonic-cosmological simulation suites. To evaluate the dependence of orbit quality on the temporal cadence of the available snapshots, we re-integrate our sample of selected stars while sampling the potential model less frequently compared to our fiducial time interval of 25 Myr for approximately 4 Gyr (to present day). While one can integrate each orbit for a fixed number of orbital periods, which would decrease the overall errors, our fixed time approach is more representative of integration needs for a statistical ensemble of orbits.

Fig.~\ref{fig:intervals} plots the fraction of stars with errors below specific error thresholds (different colors) in recovered properties: total angular momentum ($J_{\text{tot}}$), total energy ($E_{\text{tot}}$), and the positional error metric ($\frac{\Delta r}{r}$ from eq.~\ref{eq:error_metric}) as a function of sampling interval. An eightfold increase in the snapshot spacing (to 200 Myr) does not significantly affect the recovery errors for $J_{\text{tot}}$ and $E_{\text{tot}}$. The fraction of stars with well-recovered orbits only starts to decrease significantly at a sampling interval of 500 Myr, a twentyfold increase from our fiducial sampling. The positional error metric ($\frac{\Delta r}{r}$) is more sensitive to the sampling interval. Stars show a noticeable deterioration in positional accuracy at a sampling interval of 100 Myr.

 This finding is consistent with DM-only results from \citet{sanders2020models}, although we caution that it depends strongly on the period of the orbits to be reconstructed. Our sample population consists of halo-like orbits with relatively long periods, while dynamical times in the disk and bulge of our simulations can be well below our snapshot frequency; we would expect only orbit-averaged properties to be stable in that case. Indeed, part of the reason for higher fractional errors in position is due to our choice of integrating over a fixed time—which could be a factor of 100 in orbital periods for different stars. For instance, a star in the outer halo with a period of 2 Gyr is only integrated for 2 periods, while a star in the inner region with a period of 0.2 Gyr is integrated for 20 periods. Moreover, due to expected density profiles, there are more stars with shorter periods. In fact, 60\% of stars have orbital periods $\leq$ 0.8 Gyr (see Fig.~\ref{fig:Tfail}).

\section{Simulating stream formation} \label{sec:apps}

\begin{figure*}
\includegraphics[width=\textwidth,]{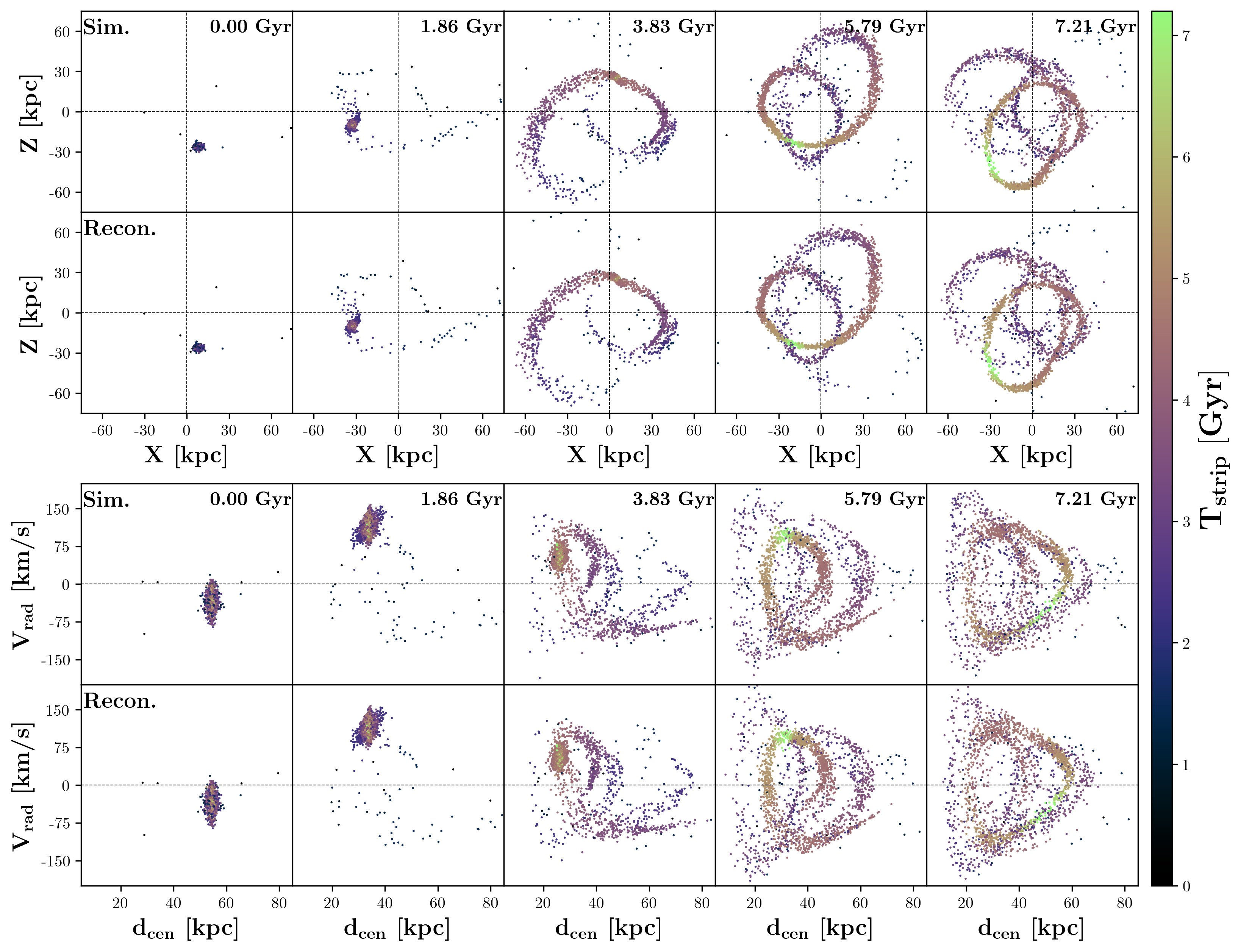}
\caption{{\bf Recovering tidal evolution in the potential model.} Tidal disruption of a dwarf satellite with $\textrm{M}_\star = 10^7 \Msol{}$ identified in \citep{panithanpaisal2021galaxy} in the live simulation (first and third rows) and in the time-evolving potential model (second and fourth rows) over a period of 7.2 Gyr of evolution (arranged in columns from left to right), color-coded by a star's stripping times. The top two rows show the stream's distribution in the XZ plane at different times. The bottom two rows present the phase-space distribution (radial velocity versus distance from the center). Reconstructed streams demonstrate excellent recovery of the stream's structure and kinematics.  \label{fig:m12i_stream}}
\end{figure*}

In this section, we demonstrate an application of the TEMP models: forming realistic stellar streams. Stellar streams are formed when stars in a globular cluster or a dwarf galaxy are tidally stripped by a more massive host galaxy \citep[e.g.,][]{toomre1972galactic, helmi1999building}, making their structure and evolution highly sensitive to the host galaxy's potential and mass profiles \citep[e.g.,][]{johnston1999tidal, law2010sagittarius, koposov2010constraining, bonaca2014milky}. Therefore, a successful potential model of the host should accurately replicate the observed evolution and structure of stellar streams.

We use our TEMP model to reconstruct an example stellar stream with $\textrm{M}_\star = 10^7$ \Msol{} at present day formed by tidal disruption of a dwarf satellite {of similar mass} ($\textrm{M}_\star = 10^7$ \Msol{}), identified by \cite{panithanpaisal2021galaxy} in \mi{}. {The progenitor dwarf galaxy and subsequently the stream form self-consistently inside our cosmological zoom-in box \mi{}. This setup includes all relevant physical processes that influence stream formation and evolution, such as internal dynamical evolution, galactic tidal forces, and dynamical friction, all modeled at the resolution allowed by our particle fidelity. The resulting stream consists of around 2000 star particles.} We begin tracking tidal disruption at the ``stream formation time'' for this progenitor galaxy, defined as in  \cite{panithanpaisal2021galaxy}, which marks the first instance when the tidal deformation of the progenitor--measured by its principal axis ratio of the moment-of-inertia tensor--exceeds a threshold value. {Since the moment-of-inertia tensor represents the overall shape of the progenitor, our disruption metric effectively captures shape deformation. Consequently, a small fraction of star particles ($\leq 0.5\%$) may already be unbound by the time we define stream formation. However, we include them for consistency in the analysis.} For our example progenitor this occurs $6.5$ Gyr after the Big Bang, which we will call $T'=0$ Gyr, representing the start of orbit integration. We assign unique stripping times ($T_\textrm{strip}$) to each star associated with the satellite by tracking when it crosses twice the time-evolving virial radius of the progenitor satellite: $|\vec{r}_* - \vec{r}_{\mathrm{sat}}| > 2R_{200m}^\mathrm{prog}(t)$. Once a star reaches this distance, we consider it to be unaffected by the self-gravity of its progenitor galaxy, which is not included in the TEMP model. We then begin orbit integration of the stripped star in the host potential up to the present day. The duration of orbit integration differs for each star, but does not exceed 7.2 Gyr for those stars that are assigned as stripped at or before $T'=0$ Gyr. Once we have our potential models, these integrations are computationally inexpensive, allowing us to resimulate tidal stream formation at high resolution in a few minutes, without the cost of re-running the entire simulation.

Fig.~\ref{fig:m12i_stream} shows the XZ plane (top half) and phase-space (radial velocity versus distance from the center, bottom half) distribution and temporal evolution (arranged in columns--increasing time from left to right) of this dwarf satellite in both the live simulation (first and third row) and TEMP model (second and fourth row) over a period of 7.2 Gyr to present day. The stars are color-coded by their stripping times from the progenitor starting from $T'=0$ Gyr. Simulated stream structure (position-space; top 2 rows) and kinematics (phase-space; bottom 2 rows) align closely with the real stream, demonstrating the model's effectiveness in modeling tidal disruption and evolution. However, minor discrepancies in both position-space and phase-space are noted after 5.8 Gyr near the outer tails (further than 60 kpc from the center). Similar features are noted for other streams across the \textit{Latte} suite \citep{bregou2023effects}.

\section{Summary and discussion} \label{sec:disc_conc}

In this paper, we assess the effectiveness of the time-evolving potential (TEMP) model, fit to a zoomed baryonic-cosmological simulation of a MW-mass galaxy from the FIRE-2 suite \citep{Hopkins_2018, wetzel2023public}, introduced by \citet{arora2022stability} in recovering particle forces and halo star orbits for a $\sim4$ Gyr integration. The TEMP model incorporates a spherical harmonic expansion for the halo and azimuthal harmonic expansion for the disk, with a maximum pole order of 4.

We recover individual particle forces to high accuracy (Sec.~\ref{sec:recon_forces_total}), with 68\% of errors within 1\% and about 95\% of particles exhibiting less than 4\% error in force reconstruction at the present day (Fig.~\ref{fig:Fstd_n4n4}). We observe negligible improvement in recovered forces with increasing pole orders beyond 4, as evident from the relatively constant means and standard deviations of the force error distribution (Table~\ref{tab:force_res_order_l}). Similar minor improvements for reconstructed orbits were noted for pole orders beyond 4 in \citet{sanders2020models} and other previous works, which used DM-only simulations. 

The largest force errors produced by the model are localized near the galactic center, and are due to complex small-scale structures, such as spiral arms in the baryonic disk components. Additionally, errors arise from interactions with a few subhalos that are not resolved in our smooth density field, yet still massive enough to affect orbits. These errors can bias predictions of orbits for stars that spend the majority of their time in the inner regions ($\leq 10-15$ kpc, $\sim$ $1-1.5\times$ radius enclosing 90\% stellar mass). However, they have minimal impact on halo-like orbits, as these stars move very quickly near pericenter. Consequently, the increased error in acceleration during the shorter time spent in the inner region has a negligible effect on their overall velocities. 

We also imposed various symmetry conditions on the BFE-based potential model at present day and evaluated the recovered forces. We show that for halo particles, imposing even the most restrictive symmetries, such as axisymmetry, on the disk potential still yields accurate force reconstructions. For disk orbits, however, it is crucial to accurately model the disk, while simpler halo models can suffice (see Fig.~\ref{fig:Fstd_model_violin}, and~\ref{fig:Fstd_all_violin}). Symmetry assumptions can introduce significant biases if applied without regard for orbit type, particularly in computing integrals of motion for disk orbits.

Overall, we achieve high fidelity in orbit recovery using the TEMP model, demonstrating its effectiveness statistically (see Fig.~\ref{fig:phase_space_error}). However, the fidelity in individual orbit reconstruction depends on the specific question being addressed. For example, accurately determining individual orbits in terms of exact positions, velocities, and times is most challenging. Using an error metric based on recovering exact positions, we show that 3D positions can still be recovered to within 10\% accuracy over 2-3 orbital periods (see Fig.~\ref{fig:delta_r_evolv_hist}), though errors are higher for orbits closer to the galactic center. Notably, 70\% of orbits have total positional errors below 100\% after multiple periods (1--20) of integration (see Fig.~\ref{fig:delta_r_final}), indicating that while some errors exist, a substantial portion of the orbits can maintain reasonably accurate positional information. Therefore, while single orbits may not hold much meaning due to inherent uncertainties, a statistical ensemble of orbits can still be effectively utilized to understand the overall dynamical behavior of the system (see Fig.~\ref{fig:m12i_stream}).

This method can be particularly successful for studying halo orbits and their precise positions, such as those of the Sagittarius dwarf satellite and its tidal stream, where the larger spatial extent and statistical sample allow for more accurate orbit recovery \citep{johnston1995disruption, ibata1996kinematics, jiang2000orbit}. Conversely, orbits within the inner regions, such as those associated with the MW bar and disk, will be highly biased.

Integrals of motion, such as total energy and angular momentum, can be recovered with much higher accuracy. We find that 87\% of orbits have recovered energy errors below 100\%, and 98\% of orbits have recovered angular momentum errors under 100\% (see Fig.~\ref{fig:energy_mom}). These quantities change appreciably over time for many stars, indicating that our model accurately predicts their variations as the galaxy evolves. Moreover, these integrals are recovered to high fidelity even with a larger potential sampling interval of 200 Myr compared to our fiducial snapshot spacing of 25 Myr.   

Additionally, orbital properties that directly depend on energy and angular momentum, such as the shape of the orbits—including pericenter and apocenter distances—are well reconstructed to within 10\% accuracy even after multiple passages (see Fig.~\ref{fig:energy_mom} and Table~\ref{tab:error_integ_motion}), with apocenter distances being more accurate than pericenter distances. Interestingly, the instances where total angular momentum errors exceed 100\% do not show any significant dependence on the orbital period or the average distance from the galactic center (see Fig.~\ref{fig:Tfail}), given that total angular momentum is approximately conserved for adiabatic changes in the potential \citep{binney2008dynamics}. This implies that stream- and potential-modeling techniques based on integrals of motion should be more robust to biases induced by phase-space errors than those that compare model predictions for positions and velocities.

The high accuracy in recovering approximate integrals of motion also leads to exceptionally reliable modeling of orbital planes. This reliability is crucial for studying the planes of satellite galaxies, as dwarf galaxies in the MW appear to lie in a plane that approximately follows the Magellanic Stream, a phenomenon observed in other galaxies as well \citep{lynden1976dwarf, pawlowski2018planes}. Accurate modeling of orbital planes enhances our understanding of these satellite planes, which have important implications for the formation and evolution of galaxies and their satellite systems, as well as for dark matter \citep{sales2022baryonic, sales2023planes, sawala2023milky}.

For these accuracy statistics, we have focused on a simulation that had no massive mergers ($M_\textrm{sat} \leq 2 \times 10^{10}$ \Msol{}) during the integration time, so all of our errors can be treated as lower bounds. However, an obvious question arises: ``How big of a merger can this model accommodate before it breaks?'' The success of these integration techniques largely depends on how well the potential model can describe the deforming halo along with the merging satellite. \citet{arora2022stability} used a metric to measure how well action-space coherence was preserved, which is slightly different from, but akin to, the conservation metrics in this paper. They found that these potential modeling techniques could effectively describe mergers with mass ratios of approximately 1:15 (total mass of halo: total mass of satellite at the time of pericentric passage), including mergers similar to the Sagittarius/SMC-mass satellites, in m12f (note Fig.~3 in \citet{donlon2024debris} showing tidal debris of reconstructed orbits to high accuracy). However, the model breaks down for highly radial, massive mergers with a mass ratio of about 1:8 with the first pericentric passage very close to the center of the host, as seen in m12w. Later, \citet{arora2023lmc} and \citet{drouplic2024prep} showed that stellar stream orbits can be reproduced fairly well with errors within $10-20\%$ for halos with mergers of LMC-like orbit and mass, with a mass ratio of roughly 1:10. Additionally, \citet{drouplic2024prep} noted that a statistical ensemble of integrated stellar streams are reproduced with 20\% errors in position space (two folds higher compared to this work) after 2 orbital periods in m12b.

The TEMP-based orbit reconstruction methods find extensive application in zoomed cosmological simulations. Foundational work focused on dark matter only simulations to examine subhalo evolution and disruption \citep{lowing2011halo}, simulate stellar streams in a smoothed self-consistent field potential to analyze morphological differences between smooth and lumpy potentials \citep{ngan2015simulating}, and explore the effects of time-dependent potentials on MW satellite orbits \citep{sanders2020models}.

We expand the utility of these techniques to a fully baryonic-cosmological zoomed simulation, demonstrating their broader applicability. The models and orbit reconstruction techniques presented here have proven effective in various applications, including studying the time-evolution of stellar streams in action-space \citep{arora2022stability}, and position-space \citep{bregou2023effects, drouplic2024prep}, and deriving orbital parameters of disrupting satellites \citep{horta2023observable}, stellar streams \citep{panithanpaisal2022constraining, shipp2023streams}, and stars in the disk \citep{ansar2023bar}. Additionally, \citet{arora2023lmc} applied these models to inject and integrate synthetic stream orbits in a halo undergoing a merger with an LMC-mass satellite, while \citet{donlon2024debris} used them to increase the particle resolution of merging dwarf galaxies. Furthermore, \citet{drouplic2024prep} integrated known progenitors of dwarf galaxy streams in \textit{Latte} suite to measure the impact of the LMC on their orbits. These applications underscore the versatility and robustness of time-evolving BFE-based models in capturing complex dynamics within cosmological contexts, providing crucial insights into the formation and evolution of galaxies and their substructures.

\begin{acknowledgments}
AA and RES acknowledge support from the Research Corporation through the Scialog Fellows program on Time Domain Astronomy, from NSF grants AST-2007232 and AST-2307787, and from NASA grant 19-ATP19-0068. RES is supported in part by a Sloan Fellowship. AW received support from: NSF via CAREER award AST-2045928 and grant AST-2107772; NASA ATP grant 80NSSC20K0513; HST grant GO-16273 from STScI. SL acknowledges support from NSF grant AST-2109234 and HST grant AR-16624 from STScI. ECC acknowledges support for this work provided by NASA through the NASA Hubble Fellowship Program grant HST-HF2-51502 awarded by the Space Telescope Science Institute, which is operated by the Association of Universities for Research in Astronomy, Inc., for NASA, under contract NAS5-26555. NS is supported by an NSF Astronomy and Astrophysics Postdoctoral Fellowship under award AST-2303841.

This research is part of the Frontera computing project at the Texas Advanced Computing Center (TACC). Frontera is made possible by National Science Foundation award OAC-1818253. Simulations in this project were run using Early Science Allocation 1923870, and analyzed using computing resources supported by the Scientific Computing Core at the Flatiron Institute. This work used additional computational resources of the University of Texas at Austin and TACC, the NASA Advanced Supercomputing (NAS) Division and the NASA Center for Climate Simulation (NCCS), and the Extreme Science and Engineering Discovery Environment (XSEDE), which is supported by National Science Foundation grant number OCI-1053575.

FIRE-2 simulations are publicly available \citep{wetzel2023public} at \url{http://flathub.flatironinstitute.org/fire}. Additional FIRE simulation data is available at \url{https://fire.northwestern.edu/data}. A public version of the \textsc{Gizmo} code is available at \url{http://www.tapir.caltech.edu/~phopkins/Site/GIZMO.html}.

\end{acknowledgments}

\vspace{5mm}

\software{IPython \citep{ipython}, Matplotlib \citep{matplotlib}, Numpy \citep{numpy}, {Gizmo Analysis} \citep{2020ascl.soft02015W}, Agama \citep{vasiliev2019agama}, Rockstar \citep{behroozi2012rockstar}, {Halo Analysis} \citep{2020ascl.soft02014W}, CMasher \citep{cmasher}.}

\bibliography{reference}{}
\bibliographystyle{aasjournal}

\appendix

\newcommand{\ul}{\bf}
\newcommand{\half}{\frac{1}{2}}
\newcommand{\ph}[1]{\ensuremath{\phantom{#1}}}

\newcommand{\accelg}{\mathfrak{g}}
\newcommand{\fieldg}{g}
\newcommand{\metricg}{\mathsf{g}}

\newcommand{\LComega}{\mathsf{\omega}}
\newcommand{\Komega}{\omega}
\newcommand{\rotOmega}{\mathsf{\Omega}}
\newcommand{\fieldOmega}{\Omega}

\newcommand{\gandH}{\ensuremath{{\ul \fieldg}} and \ensuremath{{\ul\fieldOmega}}}
\newcommand{\MandN}{\ensuremath{\mathsf{M}} and \ensuremath{\mathsf{N}}}
\newcommand{\gHMN}{\ensuremath{{\ul \fieldg}}, \ensuremath{{\ul \fieldOmega}}, \MandN}

\newcommand{\order}{\mathcal{O}}

\section{Non-inertial forces on a cosmological background} \label{app:NIF_exp_back}

\subsection{Definitions}

We wish to compute the contributions to the equation of motion that arise by transforming from the global inertial frame of the simulation volume into a time-varying galactocentric frame in which the potential $\Phi$ of the simulated galaxy is modeled. This transformation can be written as a \textbf{Lorentz transformation}, $\Lambda(\vec{u}(t),\vec{\theta}(t))$, where $\vec{u}(t)$ is the instantaneous velocity of the galaxy's center of mass in the global inertial frame of the simulation box, and $\theta = \hat{n}(t)\theta(t)$ is the rotation from the coordinates in the simulation box to the instantaneous frame aligned with the galactic disk, written as a rotation by angle $\theta$ around the normal unit vector $\hat{n}$. The rotation matrix is related by
\begin{equation}\label{Rodriquez}
  R^I_{\ \,J}=\delta^I_{\ \,J}\cos\theta+n^In_J(1-\cos\theta)
    +\epsilon^I_{\ \,JK}n^K\sin\theta\ ,
\end{equation}
where we have used Einstein summation convention (repeated indices are summed). All these quantities are functions of the time $t$ (i.e. time in the simulation box).

This transformation takes place on an expanding cosmological background in which the galaxy is evolving. In conformal Newtonian gauge (useful for this situation) the metric is
\begin{equation}
ds^2 = a^2(\eta) \left\{ -\left[1+2\Phi(\vec{x})\right] d\eta^2 + \left[1-2\Phi(\vec{x}) \right] d\vec{x}^2 \right\}, 
\end{equation}
where $\eta$ is the conformal time (related to the proper time $\tau$ by $a d\eta = d\tau$), $a(\eta)$ is the scale factor and $\Phi(\vec{x})$ is the galactic potential defined in the comoving coordinates of the frame of the simulation box. Formally $\Phi$ is also a function of time but for the purposes of the Lorentz transformation we can treat it as instantaneously static; we ignore the time-dependence of $\Phi$ in what follows.

The transformation from the global inertial coordinate frame of the simulation box to the locally orthonormal galactocentric frame of the model can be described by an object called the \textbf{tetrad}, which relates the  basis vectors in the two frames:
\begin{equation}\label{tetrad}
  {\bf e}_\mu(x)=e^A_{\,\mu}(x)\,{\bf e}_A(x)\ .
\end{equation}
The Greek indices hereafter indicate components in the coordinate (simulation-box) basis, while the capital Latin letter indices indicate components in the orthonormal basis where the potential is modeled and the orbit integrations are to be carried out. The tetrad enables one to translate coordinate vector components into
orthonormal vector components, for example
\begin{equation}\label{tetrad1}
  p^A=e^A_{\,\mu}\,p^\mu\ .
\end{equation}
The reverse transformation is described by the inverse tetrad $\tilde e^\nu_{\,B}$, defined
by
\begin{equation}\label{invtetrad}
  e^A_{\,\mu}\,\tilde e^\mu_B=\delta^A_{\,B}\ ,\ \
  \tilde e^\mu_A\,e^A_{\,\nu}=\delta^\mu_{\,\nu}\ ,
\end{equation}
so that e.g. $p^\mu=\tilde e^\mu_A\,p^A$. The metric itself is formed
from the tetrad:
\begin{equation}\label{metric1}
  \metricg_{\mu\nu}=\eta_{AB}\,e^A_{\,\mu}\,e^B_{\,\nu}\ .
\end{equation}
If the potential is weak ($\Phi/c^2 \ll 1$) then we can factor the metric into the tetrad:
\begin{equation}
e^A_{\,\mu} = a(\eta)\left( \begin{array}{cccc}
(1+\Phi) & 0 & 0 & 0 \\
0 & (1-\Phi) & 0 & 0 \\
0 & 0 & (1-\Phi) & 0  \\
0 & 0& 0 & (1-\Phi) \\ 
\end{array} \right)
\end{equation}
and its inverse:
\begin{equation}
e^A_{\,\mu} = \frac{1}{a(\eta)}\left( \begin{array}{cccc}
(1-\Phi) & 0 & 0 & 0 \\
0 & (1+\Phi) & 0 & 0 \\
0 & 0 & (1+ \Phi) & 0  \\
0 & 0& 0 & (1+\Phi) \\ 
\end{array} \right)\ .
\end{equation}

\subsection{Equation of motion}

The equation of motion (equivalent to Newton's second law, $\vec{F}=d\vec{p}/dt$) of a particle in this system is the \textbf{geodesic equation}, 
\begin{equation}\label{geodesic2}
  \frac{dp^A}{d\tau}=-\frac{1}{m} \LComega^A_{\mu B}p^\mu p^B\ ,
\end{equation}
where $\tau$ is the proper time and the {\bf Lorentz connection} (also known as spin connection or
Ricci rotation coefficients) is computed from the tetrad using a version of the first Cartan structure equation (a fundamental equation of manifold geometry):
\begin{equation}
\label{cartan1Solved}
\omega^{A}_{\ph{A}B\mu} = 
e^{\nu}_{B}\partial_{[\nu}e^{A}_{\mu]} +
\eta^{AC}\eta_{BD}e^{\alpha}_{C}\partial_{[\mu}e^{D}_{\alpha]} +
\eta^{AC}\eta_{DF}e^{\alpha}_{C}e^{\nu}_{B}e^{F}_{\mu}\partial_{[\nu}e^{D}_{\alpha]}
\end{equation}
The brackets around lower indices are shorthand for the antisymmetric part:
\begin{equation}
\label{antisymmDefined}
\partial_{[\nu}e^{A}_{\mu]} \equiv \half \left( \partial_{\nu} e^{A}_{\mu} - \partial_{\mu} e^A_{\nu}\right)
\end{equation}
In this version of the equation of motion one sees two types of momenta (velocities): the coordinate momentum $p^\mu \equiv m v^\mu$, which is related to the comoving velocity in the simulation frame by a factor of the inertial mass $m$, and the momentum in the galactocentric frame, $p^B \equiv E v^B$ which is instead proportional to the energy $E$. Just like positions, the two are related by the tetrad: $p^A=me^A_{\,\mu}(dx^\mu/d\tau)$. The tetrad also relates the coordinate time $\tau$ on the left-hand side of Equation \ref{geodesic2} and the time $t$ in the galactocentric frame; these are related by the tetrad such that $t =  a(\tau)(1-\Phi)\tau$. For $\Phi/c^2 \ll 1$ this reduces to $t=a(\tau) \tau$.

By using the identities above, one can rearrange the equation to give the equation of motion for the 3-momentum, \emph{almost} entirely in terms of quantities defined in the galactocentric frame:
\begin{equation}
\frac{dp^I}{dt} = - E \omega^I_{\mu B} v^\mu v^B.
\end{equation}
The subscripts $IJK\ldots$ are used to indicate spatial components of the four-vectors (i.e. normal three-dimensional positions and velocities) in the orthonormal frame (i.e. one that has orthogonal unit vectors, like global Cartesian coordinates), while $ijk\ldots$ indicate 3D positions and velocities in the coordinate frame (where the unit vectors point along the different coordinate directions and might not be orthogonal, since mass ``bends'' the coordinate grid according to the equivalence principle). The inertia of a particle is given by its energy $E$, which since our velocities are all slow compared to $c$ is approximately equivalent to the rest mass $m$ and is therefore constant in time. Pulling this factor out on both sides of the equation above leaves us with
\begin{equation}
\frac{dv^I}{dt} = - \omega^I_{\mu B} v^\mu v^B.
\end{equation}

\subsection{Fields and transformations}
The terms in the equation of motion (Equation \ref{geodesic2}) can be decomposed into a form similar to the equation of motion for a particle in an electromagnetic field. This is done by separating the components of the connection by the number of powers of $v$ they contain and whether they are symmetric or antisymmetric:
\begin{equation}\label{geodesic4}
  \frac{dv^I}{dt}=g^I+M^I_{\ \,j}v^j+\epsilon^I_{\ \,KL}
    (\fieldOmega^K+N^K_{\ \,j}v^j)v^L,
\end{equation}
where we have defined
\begin{equation}\label{Gdecomp}
  \LComega^I_{t0}=\LComega^0_{tI}=-g^I\ ,\ \
  \LComega^I_{tJ}=\epsilon^I_{\ \,JK}\fieldOmega^K\ ,\ \
  \LComega^I_{j0}=\LComega^0_{jI}=-M^I_{\ \,j}\ ,\ \
  \LComega^I_{jK}=\epsilon^I_{\ \,KL}N^L_{\ \,j}\ .
\end{equation}
The object $\epsilon^I_{\ \,JK}$ is the totally antisymmetric matrix or Levi-Civita pseudotensor as found in the component definition of the cross product,
\begin{equation}
({\bf A} \times {\bf B})^I \equiv \epsilon^I_{\ \,JK} A^J B^K.
\end{equation}

This formalism is useful for our purposes since these fields transform in particular ways under the time-varying Lorentz transformation we need to arrive in the galactocentric frame at each timestep. This is because although the Lorentz connection is a tensor (specifically a one-form) under arbitrary coordinate transformations,\footnote{this is an extension of the idea that a vector always points in the same direction no matter what coordinate system you use to describe its components} it's a tensor only for global Lorentz transformations, not for time-varying or spatially-varying ones:
\begin{equation}\label{LTomega}
  \LComega^A_{\mu B}\to \bar\LComega^A_{\mu B}=\Lambda^A_{\,C}\,
    (\Lambda^{-1})^D_{\,B}\,\LComega^C_{\mu D}-(\Lambda^{-1})^C_{\,B}
    \,\partial_\mu\Lambda^A_{\,C}\ .
\end{equation}
The terms that are proportional to partial derivatives of $\Lambda$ are nonzero if the boost velocity $\vec{u}$ or the rotation angle $\vec{\theta}$ is a function of time or space.

In particular, since the transformation (both the velocity shift and the rotation) varies in time but not space, the following table gives the transformation rules that lead to the  equation of motion in the galactocentric frame, in the limit that $u\ll c$:

\begin{table}[h]
\caption{Lorentz transformations of gravitational fields.}
\label{bigtableofboosts}
\begin{center}
\begin{tabular}{|rlll|}
\hline
\multicolumn{4}{|c|}{Transformation of fields under velocity transformation (boost) by ${\bf u}(t)$ and rotation by $\theta^i \equiv \theta n^i$} \\
\hline \hline
&Non-time-varying terms& Terms if ${\bf u}(t)$  & Terms if ${\bf \theta}(t)$ \\
\hline
&&&\\
  $\bar{\fieldg}^I=$& 
  $  R^I_{\ \,J}{\bf \fieldg}^J - \left({\bf u}\times{\bf\fieldOmega} \right)^I $&
  $-\partial_t u^I $ & 
  0 \\
&&&\\
  $\bar{\fieldOmega}^I = $& 
  $R^I_{\ \,J}{\bf\fieldOmega}^J + \left({\bf u}\times {\bf \fieldg} \right)^I$&
  $-\half \left({\bf u} \times \partial_{t}{\bf u} \right)^I$ & 
  $-n^I\partial_t\theta  - \sin{\theta} \partial_{t}n^I + \left(1-\cos{\theta}\right)
  \epsilon^I_{\ \,JK} n^J \partial_{t}n^K$\\
&&&\\
  $\bar{M}^I_{\ \,k}=$ & $ R^I_{\ \,J}M^J_{\ \,k}
  -\epsilon^I_{\ \,KL}u^K N^L_{\ \,k}$ &
  $ 0 $&
  $ 0 $\\
&&&\\
  $\bar{N}^I_{\ \,k}=$& $ R^I_{\ \,J} N^K_{\ \,k} +
  \epsilon^I_{\ \,KL} u^KM^L_{\ \,k} $&
  $ 0 $ &
  $ 0 $ \\
&&&\\
\hline
\end{tabular}
\end{center}
\end{table}

This formalism thus allows us to account for \emph{both} the transformation to the galactocentric frame varying with time \emph{and} the expanding cosmological background.

\subsection{Equation of motion in the galactocentric frame}

We can use Equation \ref{cartan1Solved} and the definitions in Equations \ref{Gdecomp} to calculate the fields for our metric and tetrad. Then we can use the table to find the equation of motion in the frame of integration.

The fields in the simulation frame are:
\begin{eqnarray}
g^I &=& -\partial^I \Phi\\
\Omega^I &=& 0 \\
M^I_{\ \,j} &=& -(1-2\Phi)\frac{\dot{a}}{a} \delta^I_{\ \,j}\\
N^I_{\ \,j} &=& 0 
\end{eqnarray}

When transformed to the galactocentric frame, we find
\begin{eqnarray}
\bar{g}^I &=& -R^I_{\ \, J}\partial^J \Phi - 2u^I(\vec{u}\cdot \vec{\nabla} \Phi) - \partial_t u^I \\
\bar{\Omega}^I &=& (\vec{u}\times \vec{\nabla} \Phi)^I - \half(\vec{u}\times \partial_t \vec{u} )^I -n^I\partial_t\theta  - \sin{\theta} \partial_{t}n^I + \left(1-\cos{\theta}\right)  \epsilon^I_{\ \,JK} n^J \partial_{t}n^K \\
\bar{M}^I_{\ \,j} &=& -(1-2\Phi)\frac{\dot{a}}{a} R^I_{\ \,K}\delta^K_{\ \,j} \\
\bar{N}^I_{\ \,j} &=&  (1-2\Phi)\frac{\dot{a}}{a}\epsilon^I_{\ \,JK} \delta^J_j u^K
\end{eqnarray}

Inserting these terms into the equation of motion gives us
\begin{eqnarray}
  \frac{dv^I}{dt} &=& -\partial^I \Phi - 2u^I(\vec{u}\cdot \vec{\nabla} \Phi) - \partial_t u^I \nonumber \\
  && - \left[ (\vec{u} \times \vec{\nabla} \Phi) \times \vec{v} \right]^I - \half\left[(\vec{u}\times \partial_t \vec{u} ) \times \vec{v} \right]^I \nonumber \\
  &&-\partial_t \theta(\hat{n}  \times \vec{v})^I  - \sin{\theta} (\partial_{t}\hat{n} \times \vec{v})^I + \left(1-\cos{\theta}\right)\left[(\hat{n} \times \partial_t \hat{n}) \times \vec{v}\right] \nonumber \\
&& -(1-2\Phi)\frac{\dot{a}}{a} v^I + (1-2\Phi)\frac{\dot{a}}{a}  \left[v^2 u^I -v^I (\vec{u} \cdot \vec{v}) \right].
  \end{eqnarray}
 which includes the ``non-inertial forces''---terms from transforming from the expanding cosmological box (the simulation frame) into a local inertial frame that varies with time (the galactocentric frame). The terms should all be evaluated in the galactocentric frame. The last term on line 1 of the above equation is the ``force" from accelerating the reference frame, and the first term on line 3 is the Coriolis force from rotating the reference frame. The other terms may be fairly small to zero given that $u/c \ll 1$, $v/c \ll 1$ and $\Phi/c^2 \ll 1$. Also note the terms involving the Hubble parameter $\dot{a}/a$.

Re-introducing appropriate factors of $c$ on both sides gives
\begin{eqnarray}
  \frac{dv^I}{dt} &=& -\partial^I \Phi - \partial_t u^I \nonumber   \\
  &&-\partial_t \theta(\hat{n}  \times \vec{v})^I  - \sin{\theta} (\partial_{t}\hat{n} \times \vec{v})^I + \left(1-\cos{\theta}\right)\left[(\hat{n} \times \partial_t \hat{n}) \times \vec{v}\right] \nonumber \\
  && - 2u^I(\vec{u}\cdot \vec{\nabla} \Phi)/c^2 - \left[ (\vec{u} \times \vec{\nabla} \Phi) \times \vec{v} \right]^I/c^2 - \half\left[(\vec{u}\times \partial_t \vec{u} ) \times \vec{v} \right]^I/c^2 \nonumber \\
&& -(1-2\Phi/c^2)\frac{\dot{a}}{a} v^I + (1-2\Phi/c^2)\frac{\dot{a}}{a}  \left[(v/c)^2 u^I -v^I (\vec{u} \cdot \vec{v})/c^2 \right].
  \end{eqnarray}
  
This implies that for $v \ll c$ the EOM reduces to
\begin{eqnarray}
  \frac{dv^I}{dt} &=& -\partial^I \Phi - \partial_t u^I \nonumber -\frac{\dot{a}}{a} v^I  \\
  &&-\partial_t \theta(\hat{n}  \times \vec{v})^I  - \sin{\theta} (\partial_{t}\hat{n} \times \vec{v})^I + \left(1-\cos{\theta}\right)\left[(\hat{n} \times \partial_t \hat{n}) \times \vec{v}\right] 
\label{eq:forces_nif}
  \end{eqnarray}
where the first line is terms from acceleration and second from rotation of the galactocentric frame. In this work we fix the rotation of the galactocentric frame, so the terms on the second line vanish, leaving the terms on the first line.  Equation \ref{eq:force} is then obtained by identifying $dv^I/dt \to \vec{F}$ (the force per unit mass in the galactocentric frame), $-\partial^I \Phi \to \vec{\nabla} \Phi$ (the gravitational field in the galactocentric frame), and $\partial_t u^I \to d\vec{u}/dt$ (the peculiar velocity of the galactic center).

It is useful to consider the size of terms on the second line of this equation, induced by changing the orientation of the galactic frame in time, compared to the size of the second term on the first line, which is induced by tracking the motion of the galactic center of mass in space. It is sometimes necessary to re-orient the galactocentric frame over the course of a simulation in order to keep the azimuthal harmonic expansion coordinates lined up with the disk symmetry axis, in which case this rotation induces fictitious forces and the terms on the second line will be nonzero. Typical center-of-mass accelerations are of order 0.3 km/s/Myr \citep{arora2022stability}. In the simulation used in this work the disk plane rotates by about 20 degrees over 7 Gyr up to the present day, so $\partial_t \theta$ is $\sim 10^{-4}$ rad Myr$^{-1}$. For typical orbital velocities of $\sim 100$ km s$^{-1}$, the first term on the second line of Equation \ref{eq:forces_nif} is thus about 0.01 km s$^{-1}$ Myr$^{-1}$. However, during a merger the disk can change its orientation far more rapidly than this, which will both necessitate rotation of the galactocentric frame and increase the size of these terms.

\section{Force residuals under different symmetries across the Cartesian axes}\label{app:force_res_axis}

\begin{figure}
\centering
\includegraphics[width=\textwidth,]{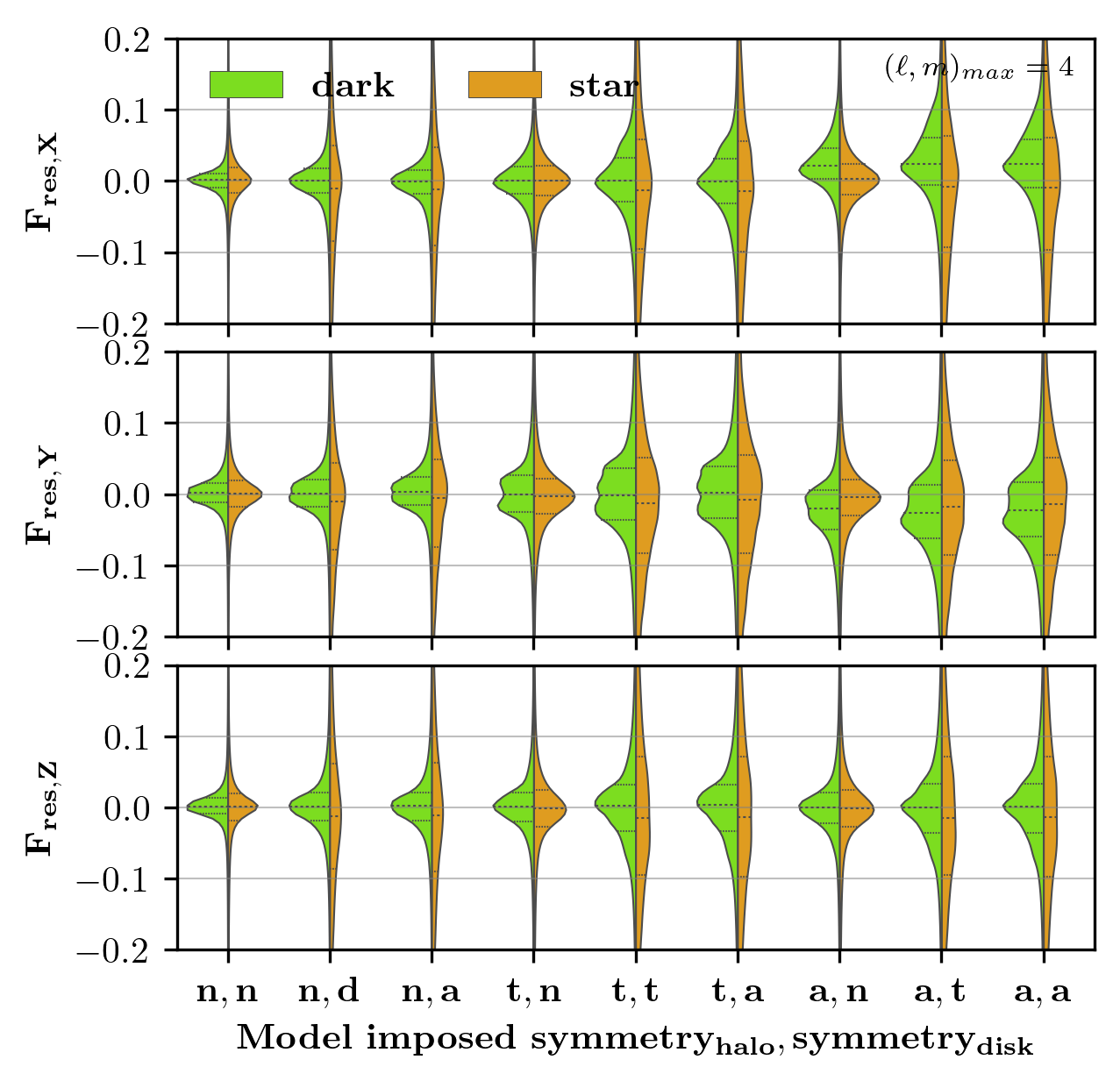}
\caption{{\bf Variation in performance with different symmetry assumptions across Cartesian axes.} Violin plot illustrating the residual distributions between the reconstructed and true forces across different Cartesian axes (first row: X axis, second row: Y axis, third row: Z axis) for DM (green) and stars (orange) within 50 kpc of the galactic center at the present day. The forces are reconstructed using the same symmetry models as in Sec.~\ref{sec:recon_forces_total}, with various symmetry conditions imposed on the halo and the disk. The notation on the x-axis indicates the symmetry conditions applied on the halo and disk respectively following the notation from Table~\ref{tab:symmetry_implications}. The short-dashed lines represent the 25th and 75th quartiles of the distributions, while the long-dashed line represents the 50th quartile. \label{fig:Fstd_all_violin}}
\end{figure}

Fig.~\ref{fig:Fstd_all_violin} presents the violin plots showcasing the residual distributions of reconstructed forces compared to true forces across different Cartesian axes for DM (green) and stars (orange) within 50 kpc of the galactic center. Each row corresponds to a specific Cartesian axis: X (top row), Y (middle row), and Z (bottom row). These distributions depict the impact of various symmetry conditions imposed on the halo and the disk, as denoted by the x-axis notation, with detailed symmetries listed in Table~\ref{tab:symmetry_implications}. The dotted lines in the plots denote the 25th and 75th quartiles, while the solid line marks the 50th quartile. 

The residual distributions across different Cartesian axes, as depicted in Fig.~\ref{fig:Fstd_all_violin}, echo the trends observed in the total force residual distributions discussed in Fig.~\ref{fig:Fstd_model_violin} (Sec.~\ref{sec:recon_forces_total}). Generally, all distributions exhibit larger standard deviations ($\sigma$) compared to the residuals of force magnitudes. However, models with constrained symmetries tend to display wider tails across all axes, along with a non-zero mean ($\mu$) along the X and Y axes. 

Notably, the residual distribution along the Z axis shows a higher degree of symmetry, with mean values close to zero and lower standard deviations for both DM and star distributions across all models. This could stem from disk alignment in the XY plane, with the disk angular momentum aligned with Z-axis. Conversely, the distribution along the Y axis exhibits a bimodal pattern, suggesting a higher likelihood of systematic biases introduced by symmetry assumptions. Interestingly, this bimodal behavior is not observed in a different simulation from the \textit{Latte} Suite \citep{wetzel2023public} (not included).
\end{document}